\documentstyle[eqsecnum,epsfig,aps]{revtex}
\newcommand{\str}{\rule{0ex}{2.7ex}}  
\newcommand{\beq}{\begin{equation}}
\newcommand{\eeq}{\end{equation}}
\newcommand{\bea}{\begin{eqnarray}}
\newcommand{\eea}{\end{eqnarray}}
\begin{document}
\draft
\preprint{SI--99--6}
%
\newbox\hdbox%
\newcount\hdrows%
\newcount\multispancount%
\newcount\ncase%
\newcount\ncols
\newcount\nrows%
\newcount\nspan%
\newcount\ntemp%
\newdimen\hdsize%
\newdimen\newhdsize%
\newdimen\parasize%
\newdimen\spreadwidth%
\newdimen\thicksize%
\newdimen\thinsize%
\newdimen\tablewidth%
\newif\ifcentertables%
\newif\ifendsize%
\newif\iffirstrow%
\newif\iftableinfo%
\newtoks\dbt%
\newtoks\hdtks%
\newtoks\savetks%
\newtoks\tableLETtokens%
\newtoks\tabletokens%
\newtoks\widthspec%
%
%
\immediate\write15{%
CP SMSG GJMSINK TEXTABLE --> TABLE MACROS V. 851121 JOB = \jobname%
}%
%
%
\tableinfotrue%
\catcode`\@=11
\def\out#1{\immediate\write16{#1}}
%
%
\def\tstrut{\vrule height3.1ex depth1.2ex width0pt}%
\def\and{\char`\&}
\def\tablerule{\noalign{\hrule height\thinsize depth0pt}}%
\thicksize=1.5pt
\thinsize=0.6pt
\def\thickrule{\noalign{\hrule height\thicksize depth0pt}}%
\def\hrulefill{\leaders\hrule\hfill}%
\def\bigrulefill{\leaders\hrule height\thicksize depth0pt \hfill}%
\def\ctr#1{\hfil\ #1\hfil}%
\def\altctr#1{\hfil #1\hfil}%
\def\vctr#1{\hfil\vbox to0pt{\vss\hbox{#1}\vss}\hfil}%
%
%
\tablewidth=-\maxdimen%
\spreadwidth=-\maxdimen%
\def\tabskipglue{0pt plus 1fil minus 1fil}%
%
%
\centertablestrue%
\def\centeredtables{%
   \centertablestrue%
}%
\def\noncenteredtables{%
   \centertablesfalse%
}%
%
%
\parasize=4in%
\long\def\para#1{
   {%
      \vtop{%
         \hsize=\parasize%
         \baselineskip14pt%
         \lineskip1pt%
         \lineskiplimit1pt%
         \noindent #1%
         \vrule width0pt depth6pt%
      }%
   }%
}%
\gdef\ARGS{########}
\gdef\headerARGS{####}
\def\@mpersand{&}
{\catcode`\|=13
\gdef\letbarzero{\let|0}
\gdef\letbartab{\def|{&&}}%
\gdef\letvbbar{\let\vb|}%
}
{\catcode`\&=4
\def\ampskip{&\omit\hfil&}
\catcode`\&=13
\let&0
\xdef\letampskip{\def&{\ampskip}}%
\gdef\letnovbamp{\let\novb&\let\tab&}
}
\def\begintable{
   \begingroup%
   \catcode`\|=13\letbartab\letvbbar%
   \catcode`\&=13\letampskip\letnovbamp%
   \def\multispan##1{
      \omit \mscount##1%
      \multiply\mscount\tw@\advance\mscount\m@ne%
      \loop\ifnum\mscount>\@ne \sp@n\repeat%
   }
   \def\|{%
      &\omit\widevline&%
   }%
   \ruledtable
}
\long\def\ruledtable#1\endtable{%
%
%
%
   \offinterlineskip
   \tabskip 0pt
   \def\widevline{\vrule width\thicksize}
   \def\endrow{\@mpersand\omit\hfil\crnorm\@mpersand}%
   \def\crthick{\@mpersand\crnorm\thickrule\@mpersand}%
   \def\crthickneg##1{\@mpersand\crnorm\thickrule
          \noalign{{\skip0=##1\vskip-\skip0}}\@mpersand}%
   \def\crnorule{\@mpersand\crnorm\@mpersand}%
   \def\crnoruleneg##1{\@mpersand\crnorm
          \noalign{{\skip0=##1\vskip-\skip0}}\@mpersand}%
   \let\nr=\crnorule
   \def\endtable{\@mpersand\crnorm\thickrule}%
   \let\crnorm=\cr
%
%
   \edef\cr{\@mpersand\crnorm\tablerule\@mpersand}%
   \def\crneg##1{\@mpersand\crnorm\tablerule
          \noalign{{\skip0=##1\vskip-\skip0}}\@mpersand}%
   \let\ctneg=\crthickneg
   \let\nrneg=\crnoruleneg
   \the\tableLETtokens
%
%
   \tabletokens={&#1}
%
%
   \countROWS\tabletokens\into\nrows%
   \countCOLS\tabletokens\into\ncols%
%
%
   \advance\ncols by -1%
   \divide\ncols by 2%
   \advance\nrows by 1%
%
%
   \iftableinfo %
      \immediate\write16{[Nrows=\the\nrows, Ncols=\the\ncols]}%
   \fi%
%
%
   \ifcentertables
      \ifhmode \par\fi
      \hbox to \hsize{
      \hss
   \else %
      \hbox{%
   \fi
      \vbox{%
         \makePREAMBLE{\the\ncols}
         \edef\next{\preamble}
         \let\preamble=\next
         \makeTABLE{\preamble}{\tabletokens}
      }
      \ifcentertables \hss}\else }\fi
   \endgroup
   \tablewidth=-\maxdimen
   \spreadwidth=-\maxdimen
}
\def\makeTABLE#1#2{
   {
   \let\ifmath0
   \let\header0
   \let\multispan0
%
%
   \ncase=0%
   \ifdim\tablewidth>-\maxdimen \ncase=1\fi%
   \ifdim\spreadwidth>-\maxdimen \ncase=2\fi%
   \relax
%
   \ifcase\ncase %
      \widthspec={}%
   \or %
      \widthspec=\expandafter{\expandafter t\expandafter o%
                 \the\tablewidth}%
   \else %
      \widthspec=\expandafter{\expandafter s\expandafter p\expandafter r%
                 \expandafter e\expandafter a\expandafter d%
                 \the\spreadwidth}%
   \fi %
   \xdef\next{
      \halign\the\widthspec{%
      #1
      \noalign{\hrule height\thicksize depth0pt}
      \the#2\endtable
%
      }
   }
   }
   \next
}
\def\makePREAMBLE#1{
   \ncols=#1
   \begingroup
   \let\ARGS=0
   \edef\xtp{\widevline\ARGS\tabskip\tabskipglue%
   &\ctr{\ARGS}\tstrut}
   \advance\ncols by -1
   \loop
      \ifnum\ncols>0 %
      \advance\ncols by -1%
      \edef\xtp{\xtp&\vrule width\thinsize\ARGS&\ctr{\ARGS}}%
   \repeat
   \xdef\preamble{\xtp&\widevline\ARGS\tabskip0pt%
   \crnorm}
   \endgroup
}
\def\countROWS#1\into#2{
   \let\countREGISTER=#2%
   \countREGISTER=0%
   \expandafter\ROWcount\the#1\endcount%
}%
\def\ROWcount{%
   \afterassignment\subROWcount\let\next= %
}%
\def\subROWcount{%
   \ifx\next\endcount %
      \let\next=\relax%
   \else%
      \ncase=0%
      \ifx\next\cr %
         \global\advance\countREGISTER by 1%
         \ncase=0%
      \fi%
      \ifx\next\endrow %
         \global\advance\countREGISTER by 1%
         \ncase=0%
      \fi%
      \ifx\next\crthick %
         \global\advance\countREGISTER by 1%
         \ncase=0%
      \fi%
      \ifx\next\crnorule %
         \global\advance\countREGISTER by 1%
         \ncase=0%
      \fi%
      \ifx\next\crthickneg %
         \global\advance\countREGISTER by 1%
         \ncase=0%
      \fi%
      \ifx\next\crnoruleneg %
         \global\advance\countREGISTER by 1%
         \ncase=0%
      \fi%
      \ifx\next\crneg %
         \global\advance\countREGISTER by 1%
         \ncase=0%
      \fi%
      \ifx\next\header %
         \ncase=1%
      \fi%
      \relax%
      \ifcase\ncase %
         \let\next\ROWcount%
      \or %
         \let\next\argROWskip%
      \else %
      \fi%
   \fi%
   \next%
}
\def\counthdROWS#1\into#2{%
\dvr{10}%
   \let\countREGISTER=#2%
   \countREGISTER=0%
\dvr{11}%
\dvr{13}%
   \expandafter\hdROWcount\the#1\endcount%
\dvr{12}%
}%
\def\hdROWcount{%
   \afterassignment\subhdROWcount\let\next= %
}%
\def\subhdROWcount{%
   \ifx\next\endcount %
      \let\next=\relax%
   \else%
      \ncase=0%
      \ifx\next\cr %
         \global\advance\countREGISTER by 1%
         \ncase=0%
      \fi%
      \ifx\next\endrow %
         \global\advance\countREGISTER by 1%
         \ncase=0%
      \fi%
      \ifx\next\crthick %
         \global\advance\countREGISTER by 1%
         \ncase=0%
      \fi%
      \ifx\next\crnorule %
         \global\advance\countREGISTER by 1%
         \ncase=0%
      \fi%
      \ifx\next\header %
         \ncase=1%
      \fi%
\relax%
      \ifcase\ncase %
         \let\next\hdROWcount%
      \or%
         \let\next\arghdROWskip%
      \else %
      \fi%
   \fi%
   \next%
}%
{\catcode`\|=13\letbartab
\gdef\countCOLS#1\into#2{%
   \let\countREGISTER=#2%
   \global\countREGISTER=0%
   \global\multispancount=0%
   \global\firstrowtrue
   \expandafter\COLcount\the#1\endcount%
   \global\advance\countREGISTER by 3%
   \global\advance\countREGISTER by -\multispancount
}%
\gdef\COLcount{%
   \afterassignment\subCOLcount\let\next= %
}%
{\catcode`\&=13%
\gdef\subCOLcount{%
   \ifx\next\endcount %
      \let\next=\relax%
   \else%
      \ncase=0%
      \iffirstrow
         \ifx\next& %
            \global\advance\countREGISTER by 2%
            \ncase=0%
         \fi%
         \ifx\next\span %
            \global\advance\countREGISTER by 1%
            \ncase=0%
         \fi%
         \ifx\next| %
            \global\advance\countREGISTER by 2%
            \ncase=0%
         \fi
         \ifx\next\|
            \global\advance\countREGISTER by 2%
            \ncase=0%
         \fi
         \ifx\next\multispan
            \ncase=1%
            \global\advance\multispancount by 1%
         \fi
         \ifx\next\header
            \ncase=2%
         \fi
         \ifx\next\cr       \global\firstrowfalse \fi
         \ifx\next\endrow   \global\firstrowfalse \fi
         \ifx\next\crthick  \global\firstrowfalse \fi
         \ifx\next\crnorule \global\firstrowfalse \fi
         \ifx\next\crnoruleneg \global\firstrowfalse \fi
         \ifx\next\crthickneg  \global\firstrowfalse \fi
         \ifx\next\crneg       \global\firstrowfalse \fi
      \fi
\relax
      \ifcase\ncase %
         \let\next\COLcount%
      \or %
         \let\next\spancount%
      \or %
         \let\next\argCOLskip%
      \else %
      \fi %
   \fi%
   \next%
}%
\gdef\argROWskip#1{%
   \let\next\ROWcount \next%
}
\gdef\arghdROWskip#1{%
   \let\next\ROWcount \next%
}
\gdef\argCOLskip#1{%
   \let\next\COLcount \next%
}
}
}
\def\spancount#1{
   \nspan=#1\multiply\nspan by 2\advance\nspan by -1%
   \global\advance \countREGISTER by \nspan
   \let\next\COLcount \next}%
\def\dvr#1{\relax}%
\def\header#1{%
\dvr{1}{\let\cr=\@mpersand%
\hdtks={#1}%
\counthdROWS\hdtks\into\hdrows%
\advance\hdrows by 1%
\ifnum\hdrows=0 \hdrows=1 \fi%
\dvr{5}\makehdPREAMBLE{\the\hdrows}%
\dvr{6}\getHDdimen{#1}%
{\parindent=0pt\hsize=\hdsize{\let\ifmath0%
\xdef\next{\valign{\headerpreamble #1\crnorm}}}\dvr{7}\next\dvr{8}%
}%
}\dvr{2}}
\def\makehdPREAMBLE#1{
\dvr{3}%
\hdrows=#1
{
\let\headerARGS=0%
\let\cr=\crnorm%
\edef\xtp{\vfil\hfil\hbox{\headerARGS}\hfil\vfil}%
\advance\hdrows by -1
\loop
\ifnum\hdrows>0%
\advance\hdrows by -1%
\edef\xtp{\xtp&\vfil\hfil\hbox{\headerARGS}\hfil\vfil}%
\repeat%
\xdef\headerpreamble{\xtp\crcr}%
}
\dvr{4}}
\def\getHDdimen#1{%
\hdsize=0pt%
\getsize#1\cr\end\cr%
}
\def\getsize#1\cr{%
\endsizefalse\savetks={#1}%
\expandafter\lookend\the\savetks\cr%
\relax \ifendsize \let\next\relax \else%
\setbox\hdbox=\hbox{#1}\newhdsize=1.0\wd\hdbox%
\ifdim\newhdsize>\hdsize \hdsize=\newhdsize \fi%
\let\next\getsize \fi%
\next%
}%
\def\lookend{\afterassignment\sublookend\let\looknext= }%
\def\sublookend{\relax%
\ifx\looknext\cr %
\let\looknext\relax \else %
   \relax
   \ifx\looknext\end \global\endsizetrue \fi%
   \let\looknext=\lookend%
    \fi \looknext%
}%
%
%
\def\tablelet#1{%
   \tableLETtokens=\expandafter{\the\tableLETtokens #1}%
}%
\catcode`\@=12
%

\title{ 
	DPMJET    version  II.5        \\
   \vspace*{3mm}
        Sampling of hadron--hadron, hadron-nucleus\\ 
          and nucleus-nucleus interactions \\
          at accelerator and Cosmic Ray energies \\
	according to the two--component Dual Parton Model
	\vspace*{2mm}                     \\
	Code manual       \\
}
\author{  J.~Ranft }
\address{
 Physics Dept. Universit\"at Siegen, D--57068 Siegen, Germany,
 e--mail: Johannes.Ranft@cern.ch}
\date{\today}
 \maketitle

\maketitle


\begin{abstract}

The physics of DPMJET--II.5 is described in a companion paper
\cite{Ranft99a}.
 DPMJET samples hadron--hadron,
 hadron--nucleus, nucleus--nucleus and neutrino--nucleus interactions 
at high energies.
The two--component Dual Parton Model is used with multiple soft
chains and multiple minijets at each elementary interaction.
Particle production is realized by the fragmentation of colorless
parton--parton chains constructed from the quark content of the
interacting hadrons.
DPMJET--II.5
includes the cascading of secondaries within the target  as well
as projectile nuclei 
which is suppressed by the formation time concept.
The excitation energy of the remaining target and projectile 
nuclei is calculated 
and using  this nuclear evaporation is included into the model.
At lab energies below $3-5~GeV$ hadron--nucleus collisions 
are described
within the conventional formation zone intranuclear cascade picture;
 thus the model may be
applied down to such energies.  
It is possible to use the model up to 
primary energies of 10${}^{21}$ eV (per nucleon) in the lab. frame.

DPMJET can also be applied to neutrino nucleus collisions. It
extends the neutrino--nucleon models qel (quasi elastic neutrino
interactions) and lepto (deep inelastic neutrino nucleon
collisions)  to neutrino collisions on nuclear targets.
\end{abstract}

\vskip 30mm
{Siegen preprint SI--99--6}

 \clearpage


\section{Program Summary}
   \vspace*{20mm}
\begin{tabbing}
shiftshiftshiftshiftshiftshiftshiftshift\=                       \kill
Title of the program:    \>DPMJET--II.5  \\
   \>   \\
Computer:                \>UNIX Workstations, LINUX--PC's  \\
  \>   \\
Program language:        \>FORTRAN-77   \\
  \>   \\
Number of program lines: \>about 80,000 (in addition  the Lund
codes linked)    \\
  \>   \\
Other programs called:   \> By DPMJET--II.5 \\
~~~(included in DPMJET   \>       \\
~~~~in modified form)    \>       \\
                         \>DIAGEN \cite{Shma88}  \\
                          \>~~~Sampling of configurations for
                              nucleus-nucleus interactions \\
                         \>~~~within the Glauber formalism  \\
                         \>DECAY \cite{DECAY}  \\
                         \>~~~Sampling the decay of hadron resonances.\\
                         \>HADRIN \cite{HADRIN}    \\
                         \>~~~Sampling hadron-nucleon interactions
                              below 5 GeV.   \\
                         \>parts of DTUJET \cite{DTUJET93,DTUJET93code}    \\
                 \>~~~Sampling of minijets and multiple soft chains.  \\
                         \>parts of FLUKA \cite{FLUKA}    \\
                  \>~~~Nuclear evaporation and residual nucleus
		  module.  \\
Other programs called:  \>qel \cite{qel}   \\
                 \>~~~Quasi elastic neutrino--nucleon interaction.  \\
                         \>PYTHIA--6.1 \cite{JETSET}    \\
                         \>~~~Sampling the hadronization of
                              strings  \\
                         \>~~~according to the Lund model
                              \cite{JETSET} \\
                    \>~~~(Double precision  version of JETSET )\\
                         \>lepto--6.5 \cite{lepto}    \\
                 \>~~~Deep  inelastic neutrino--nucleon interaction.  \\
                 \>~~~Note: lepto--6.5 uses the single precision JETSET--7.4.  \\
  \>   \\
Method of solution:      \>Monte Carlo event generator.     \\
  \>   \\
\end{tabbing}

\clearpage

%
\section{Description of the program DPMJET--II.5}
The basic event generating unit of the code is the subroutine KKINC.
Each call of this routine samples one inelastic hadron--nucleus or
nucleus--nucleus interaction. Use and necessary initializations are
described first in this section; a test program provided with the program
package demonstrates the potential application.
In the following subsection we discuss important model
parameters and define their location in the code for potential user access.
The basic structure of the supplied code is described in further
subsections.
\subsection{The event generator and its initialization}
As already mentioned the code DPMJET--II.5 uses several
other programs. The initialization of the 
{\bf DECAY}\cite{DECAY} and {\bf HADRIN}\cite{HADRIN} codes
requires {\it one single call} of the subroutines
DPRIBL, DDATAR, DCHANT, DCHANH and DHADDE.
\\
Calls to PRBLM2 and JDTU initialize the multi--Pomeron sampling and the
sampling of minijets like in the DTUJET--99 code (similar to
DTUJET--93 \cite{DTUJET93,DTUJET93code}).
\\
A call to PYINIT initializes the PYTHIA--6.1 sampling of chain
decay. 
\\
Further initializations and parameter definitions are provided in the
routine DMINIT, via the named
BLOCK DATA BLKDT1 and the subroutine DEFAUL(EPN,PPN), which 
has to be called before event generation, too.
Besides other initializations this routine sets the parameters,
characterizing the type of the actual interaction, for
$\pi^+ Cu$ collisions at 200~GeV; EPN and PPN (output variables) give energy
and momentum in the lab--system.
Actual predefinition and locations of those parameters which might
be changed potentially by the user are discussed in the following
subsection.
\\
Basic information for the application of the Glauber formalism
according to the code {\bf DIAGEN}\cite{Shma88} is generated by
by the subroutine
\\
   { \centerline{SHMAKI (IP, IPZ, IT, ITZ, RPROJ, RTARG, PPN)} }
which requires the following input parameters defining the actual
interaction:
\begin{quote}
  \begin{description}
    \item[IP, IPZ:]   nucleon number/atomic number of the projectile
		      nucleus; set IP=IPZ=1 for incident hadrons;
    \item[IT, ITZ:]   nucleon number/atomic number of the target nucleus;
    \item[PPN:]        projectile momentum in GeV/c (per nucleon)
  \end{description}
\end{quote}
Since the calculations performed by SHMAKI are time consuming, in particular
for heavy target nuclei, and in general have to be repeated for each
different reaction type and energy, resp., the test program provided offers
an option to prepare a data file 'GLAUBTAR.DAT' containing the necessary
information for
hadron--nucleus and nucleus--nucleus 
interactions to be considered. This file may be generated in
a separate run of the test program using the option 'GLAUBERI' and
GLAUBERA, compare
Appendix~A.
The information from this file is read for a given projectile (projectile
nucleus) (IP,IPZ) and target nucleus~(IT,~ITZ)
by means of a subroutine call \\
{\centerline{CALL SHMAKF(IP,IPZ,IT,ITZ).} }
For the use of different target materials and or different projectiles 
in one calculation SHMAKF has
to be called subsequently with the corresponding parameters~(IP,IPZ,IT,ITZ).
The information read from the file 'GLAUBTAR.DAT' is numbered internally
by the index KKMAT=1,2,... according to the sequence of SHMAKF calls;
up to 50 materials may be stored in the standard version of the program.
If a larger number of materials is to be used the corresponding dimension
in the common /DTUMAT/ has to be increased.
(If the required data are not found in the file 'GLAUBTAR.DAT'
the execution is stopped.)
\\
After these initializations each call of the subroutine \\
{\centerline { KKINC (EPN, IT, ITZ, IP, IPZ, IPROJ, KKMAT, ITARG,
NHKKH1, IREJ)}  }
generates a single event. The input parameters not yet described have the
following meaning:
\begin{quote}
  \begin{description}
    \item[IPROJ:]   projectile type for hadron--nucleus collisions; \\
		    DPMJET--II.5 uses the naming and internal 
                    numbering conventions from the BAMJET \cite
		   {BAMJET,BAMJET1} and DECAY \cite{DECAY} codes
		     which are listed   in
		    Table~A--2 of Appendix~A.1;
    \item[ITARG:]   Target type for hadron--hadron collisions; \\
		    DPMJET--II.5 uses the naming and internal 
                    numbering conventions from the BAMJET \cite
		   {BAMJET,BAMJET1} and DECAY \cite{DECAY} codes
		     which are listed   in
		    Table~A--2 of Appendix~A.1;
    \item[KKMAT:]   controls the access of the event generator to the
		    information on the Glauber formalism: \\
		    KKMAT=0~: Glauber data expected from SHMAKI
			      calculation; \\
		    KKMAT$>$0~: Glauber data expected from the KKMAT'th call
			      of the subroutine SHMAKF.
    \item[NHKKH1:]  gives the position in the event COMMON
    HKKEVT, after which the final state particles \\
    are recorded.
    \item[IREJ:] IREJ = 1  indicates, that the event has been
    rejected\\
    IREJ = 0 indicates, that the event is fine.
  \end{description}
\end{quote}
If DPMJET--II.5 is used as event generator in a hadron cascade
code, it is practical to write an interface routine. For the use
of DPMJET--II.4 in the Cosmic Ray cascade code HEMAS--DPM
\cite{DPMBFR94,Battistoni99a} 
there
exists such an interface. DPMJET--II.4 has also been included into the
CORSIKA code of Karlruhe \cite{Corsika}.

The following commands will cause the generator to sample one inelastic
$\pi^+ Cu$ event at 200~GeV laboratory energy:
\begin{verbatim}
      CALL DMINIT(NCASES,EPN,PPN,NCOUNT,IGLAUB)

All of the following down to CALL SAMPPT is usually done within DMINIT

  ***  DECAY initialization
        CALL DDATAR
        CALL DCHANT
  *** HADRIN initialization
        CALL DHADDE
        CALL DCHANH
  *** setting default parameters
        CALL DEFAUL(EPN,PPN)
        CALL DEFAUX(EPN,PPN)
  *** JETSET initialization
        CALL LUNDIN 
  *** initialization of the random number generator supplied with DPMJET--II.5
        CALL RNDMST(12,34,56,78)
  *** initialization for the Glauber formalism by explicit calculation
  *** for the actual reaction (projectile/target/energy = pi+/Cu/200 GeV)
        IP=1
        IPZ=1
        IT=64
        ITZ=29
        PPN=200.
        CALL SHMAKI(IP,IPZ,IT,ITZ,RPROJ,RTARG,PPN)
  *** sampling of 1 event (pi+ has type IPROJ=13)
        IPROJ=13
        KKMAT=0
  *** initialization of the evaporation module
        CALL BERTTP
        CALL INCINI
  *** initialization of the unitarization
        CALL PRBLM2(CMENER)
  *** initialization of the hard scattering
        CALL JTDTU(0)
  *** initialization of the transverse momenta for soft scattering
        CALL SAMPPT(0,PT)


  *** generating one event
        CALL KKINC(EPN,IT,ITZ,IP,IPZ,IPROJ,KKMAT)}
        STOP
\end{verbatim}
To use the information from the file 'GLAUBTAR.DAT' one has to call SHMAKF
instead of SHMAKI and define KKMAT=1.

%
%
\subsection{Information on the final state particles}
During the generation of single events several entries are scored
in the common blocks /HKKEVT/ and /EXTEVT/ 
characterizing subsequent stages of
the sampling process: Information on
initial state nucleons as well as on interacting partons and constructed
parton chains, decaying
resonances and final state particles is stored into these common blocks.
It has the following structure, completely defined in Appendix~B:
\begin{verbatim}
	PARAMETER (NMXHKK=.....)
        COMMON /HKKEVT/ NHKK,NEVHKK,ISTHKK(NMXHKK),IDHKK(NMXHKK),
       &                JMOHKK(2,NMXHKK),JDAHKK(2,NMXHKK),
       &                PHKK(5,NMXHKK),VHKK(4,NMXHKK),WHKK(4,NMXHKK)
        COMMON /EXTEVT/ IDRES(NMXHKK),IDXRES(NMXHKK),NOBAM(NMXHKK),
       &                IDBAM(NMXHKK),IDCH(NMXHKK),NPOINT(10)
\end{verbatim}
The entries are characterized by their status ISTHKK and type IDHKK as well
as by the 4-momenta PHKK;
additional pointers define 'parents' and 'daughters' of the
actual entry.
Final state particles are identified by their status ISTHKK($i$)=1,
-1 or 1001. Resonances, which have decayed are available with
ISTHKK($i$)=2.
\\
The structure of this common block closely follows
the suggestions of Ref.~\cite{ZLEP,ZLEP1};
conventions for the description of the event history are
described in some detail in Appendix~B.2, followed by a sample event in
Appendix~B.3.
\\
%
%
%
\subsection{Important parameters of the code}
In this subsection we summarize physical meaning and location of those
parameters which may have a significant influence on observable quantities.

%
{\bf Sampling of x--values for partons}
 
Parton x--values for a given hadron are sampled from distributions of the
general form \\
{\centerline
	  { $ q(x) \propto x^{-\alpha} (1-x)^{\beta}, $ }  }
subject to the reqirement that their sum is equal to one.
The user has access to the $\beta$--parameters for valence quarks which
have, however, only
minor influence on the final results. They are set to the following default
values:
\\
{\centerline
   { $\beta^{nuc}_v = 3.5,$ \hspace{1cm}  $\beta^{mes}_v = 1.5 $;}}
for reasons of an optimal sampling efficiency $\beta_{sea} \equiv 0.0 $.
From standard Regge arguments the power in x is fixed to be
$\alpha^{sea}=0.5$ and $\alpha^{val}=0.5$ for sea and valence quarks,
respectively.
\\
Additionally, the x--values for partons of colliding projectile and target
hadrons are correlated by the interaction mechanism within the DTU model:
The constructed
color--neutral parton--parton systems (chains) should aquire at least some
minimum mass $M_{min}$
to allow the fragmentation into final state hadrons,
\\
{\centerline
   {    $ M^2 \simeq x^{target}  \cdot  x^{project.}\cdot s
	     \geq  M_{min}^2$;   }  }
otherwise the collision numbers sampled according to the Glauber
formalism may be strongly biased.
Therefore, lower cuts are defined for the sampling of x--values,
    $ x^{min} =  C / \sqrt{s}$,
differing for valence quarks, diquarks and sea quarks, resp.
To ensure minimum chain masses for valence--valence (v--v) systems
the following values are set by default within DPMJET--II.5:
\\
{\centerline
  { $ C^{val}_q = 1.8,$  \hspace{1cm}  $ C^{val}_{qq} = 2.0,$
			 \hspace{1cm}  $ C^{sea}_{q/\bar{q}} = 0.5 $ } }
However, since these cuts are imposed for both the projectile and the
target independently, lower (but still kinematically allowed) chain masses
tend to be suppressed at least for v--v systems. 
 All the user accessible parameters discussed so far are located in the
common block
\begin{verbatim}
      COMMON /XSEADI/ XSEACU,UNON,UNOM,UNOSEA, CVQ,CDQ,CSEA,SSMIMA,
     +                SSMIMQ,VVMTHR
\end{verbatim}
with the following assignments:
\begin{quote}
\begin{tabbing}
  $\beta^{nuc}_v$ = UNON  \hspace*{2cm}  \=
  $  \beta^{mes}_v$ = UNOM \hspace*{2cm} \=   \hfill  \\
  $C^{val}_q$ = CVQ       \>   $C^{val}_{qq}$ = CDQ
			  \>   $C^{sea}_{q/\bar{q}}$ = CSEA   \\
  $M_{ss}^{min}$ = SSMIMA \>   $(M_{ss}^{min})^2$ = SSMIMQ   \>
  $M_{vv}^{thr}$ = VVMTHR
\end{tabbing}
\end{quote}

%
{\bf Generation of transverse momenta for secondary hadrons}
 
In the model there are two sources of transverse momenta for created
secondaries. First the partons aquire an internal $p_{\perp}$ which
is taken into account in the construction of parton--parton chains.
The subsequent hadronization of the partonic chains is modelled by 
the JETSET code \cite{JETSET}
which itself assignes to the created particles
transverse momenta with respect to the jet axis.
\\
Internal transverse momenta for partons are sampled within the
subroutine SELPT (file dpmnuc3.f) according to the following distribution for
the reduced transverse energy \mbox{$E_s = E_{\perp} - m_p$},
\beq
     \frac{dn}{dE_s} \propto  E_s \exp{(-\gamma^2 E_s / 2)}
\nonumber
\eeq
with \\
{ \centerline
  {$p_\perp = \sqrt {E_s^2 + 2 E_s m_p},$ \hspace*{2cm} $m_p = 0.94$~GeV,}}
where the slope parameter $\gamma \equiv \rm{BB3}$ is defined directly in
this routine (default value BB3=6.0).
However, in the case of severe kinematical limitations for
individual partons/chains this parameter may be modified in the
sampling process. \\

%
{\bf Intranuclear cascade}
 
There are two important parameters controlling the development of the
intranuclear cascade which have been defined in the previous section: The
formation time $\tau_0$ of created secondary hadrons,
and the factor $\alpha_{mod}^F$ scaling the Fermi momenta of nucleons.
With increasing $\tau_0$ the number of cascade generations and the number of
low-energy particles will be reduced; $\alpha_{mod}^F$ provides the
possibility for a certain modification of the momentum distribution for
low--energy nucleons.
\\
The maximum number of cascade generations KTAUGE (by default
0) has to  be reset by the user, since the cascade is switched
off by the default KTAUGE=0.
\\
In the code the corresponding parameters are defined in the common blocks
\begin{verbatim}
      COMMON /TAUFO/  TAUFOR,KTAUGE,ITAUVE
      COMMON /NUCIMP/ PRMOM(5,248),TAMOM(5,248),...,
     +                PREBIN,TAEBIN,FERMOD,ETACOU
\end{verbatim}
where $\tau_0 \equiv {\rm TAUFOR}$ and
$\alpha_{mod}^F \equiv {\rm FERMOD}$.
The option ITAUVE determines whether the $p_{\perp}$--dependent
definition of the formation time (ITAUVE=1) or the constant value $\tau_0$
(ITAUVE=2) are used; by default ITAUVE=1 (compare Section 2.2).

%
%
{\bf Lund JETSET fragmentation}
 
The chain fragmentation according to the Lund model JETSET
contained within PYTHIA--6.1 \cite{JETSET}   
is customized for the DPMJET--II.5 needs by changing some of the Lund
parameters in the initialization routine LUNDIN. More parameters are
changed in the routine BAMLUN, where one JETSET fragmentation is called.
\\
In LUNDIN we prevent furthermore the weak decays of all
otherwise stable hadrons and the decay of the $\pi ^0$ by
setting the corresponding MDCY parameters equal to zero.
%

{\bf Baryon and strangeness production,}
{\bf Popcorn mechanism:}

The popcorn mechanism in 
JETSET is controlled by the parameter $p_{dB}$ (default:
$p_{dB}$ = 0.1).

\begin{verbatim}
         COMMON /POPCOR/ PDB,AJSDEF
\end{verbatim}

The parameters for the POPCORSE effect are in the COMMON /POPCCK/
and the parameters for the CASADIQU effect are in the COMMON /CASADI/.

In JETSET $p_{dB}$ = 0. is used, to switch off the popcorn
mechanism.

{\bf Sea SU(3) symmetry:}

The parameter $s^{sea}$ = SEASQ
(default: SEASQ = 0.5)  controls the s-quark content at the
sea--quark chain ends.

\begin{verbatim}
         COMMON /SEASU3/ SEASQ
\end{verbatim}

{\bf Cronin effect:}

The parameters MKCRO (default: MKCRO = 1) and
CRONCO (default: CRONCO = 0.64) control the multiple scattering
of partons within nuclear matter, and thus the Cronin effect.

\begin{verbatim}
         COMMON /CRONIN/ CRONCO,MKCRON
\end{verbatim}

MKCRON = 0 switches off the Cronin effect.

CRONCO is the parameter in the parton multiple scattering
formula.

%
\subsection{Structure of the supplied code}

The DPMJET--II.5 standard source code is 
stored in several FORTRAN files,
listed in the following:
\begin{itemize}
  \item dpm25nuc1.f, dpm25nuc2.f, dpm25nuc3.f, dpm25nuc4.f, dpm25nuc5.f,
  dpm25nuc6.f dpm25nuc7.f
     \begin{quote}
	 These modules
	 contain all subroutines needed to describe the
	 primary interaction according to the DPM
	 as defined in the previous section. This includes
         the assignment of Fermi momenta to nucleons, construction
         of chains and monitoring their decay as well as
	 the intranuclear cascade for generated secondaries.
     \end{quote}
  \item    dpm25hadri.f
     \begin{quote}
        Includes the HADRIN \cite{HADRIN} routines for event generation
        in hadron-nucleon collisions below $5~GeV$ as well as routines
        for the calculation of energy and reaction dependent
        cross sections.
     \end{quote}
  \item    dpm25nulib.f
     \begin{quote}
        Contains few standard routines usually uneffected by further
        developments of the code, including the applied
        random number generator \cite{Marsaglia}.
     \end{quote}
  \item    dpm25hist.f
     \begin{quote}
	Contains 
	 a sample histogram routine producing  line
	printer output for average multiplicities, rapidity and
	pseudorapidity distributions for generated particles.
	This serves mainly as an example on how to use the DPMJET
	events and for benchmarking the code when installing it.
     \end{quote}
  \item   dpm25diff.f
      \begin{quote}
         Contains the routines used to call single diffractive particle                  production.
      \end{quote}
  \item  pythia61.f  
      \begin{quote}
         Contains double precision version of JETSET 
	 combined with the PYTHIA--6.1 code.
       \end{quote}
  \item  dpm25evap2.f or dpm25eva.f  
      \begin{quote}
         Contains  the nuclear evaporation module. Please note,
	 that the module dpm25evap2.f and the data file NUCLEAR.BIN 
	 can not be obtained from the
	 author of DPMJET--II.5. The permission to use the 
	 evaporation module from FLUKA
	 \cite{FLUKA} has to be obtained directly from the
	 Authors of FLUKA (Dr. A.Ferrari and Dr. P.R.Sala, CERN,
	 Geneva and INFN,
	 Sezione di Milano, I--20133 Milano, Italy). The dpm25eva.f
	 module contains dummy routines, to enable to run
	 DPMJET-II without changes without the DPMEVAP module.
	 Of course dpm25eva.f performs no evaporation of the
	 residual nucleus.
       \end{quote}
  \item  dtu25lap.f or dpm25lap.f 
      \begin{quote}
         Contains  the calculation of the minijets.

	 Using the file dtu25lap.f the initialization of the 
	  minijets  is done for one energy,
	 given by the input cards. This version is used for
	 stand--alone runs of the code, where each event is
	 called for the same primary energy.

	 Using the file dpm25lap.f this initialization is done
	 for all energies. This version is suitable to insert
	 DPMJET--II.5 into hadron cascade codes, where each event
	 will be called at  a new energy.
       \end{quote}
  \item  dtu25pom.f or dpm25pom.f 
      \begin{quote}
         Contains  the calculation of multiple chain 
	 production according to
         eikonal unitarization.

	 Using the file dtu25pom.f the initialization of the two
	 component Dual Parton Model (calculation of the exclusive
	 multi Pomeron distribution) is done for one energy,
	 given by the input cards. This version is used for
	 stand--alone runs of the code, where each event is
	 called for the same primary energy. 
	 
	 Using the file dpm25pom.f this initialization is done
	 for all energies. This version is suitable to insert
	 DPMJET--II.5 into hadron cascade codes, where each event
	 will be called at  a new energy.

	 If the code is run
	 repeatedly, it will be practical, to read these
	 initialization tables, once calculated, from an
	 external file.
       \end{quote}
  \item  dpm25qelpo.f 
      \begin{quote}
	 Contains the qel code \cite{qel} 
	 transformed to double precision
	 and modified to use the Fermi momenta of the nucleons
	 from DPMJET.
       \end{quote}
  \item  leptonew.f (and jetset74ku.f) 
      \begin{quote}
	 Contains the LEPTO--6.5 code \cite{lepto}   
	 in single precision, which
	 uses the single precision jetset74.f. 
       \end{quote}
  \item  dpm25lepto.f 
      \begin{quote}
         Contains the DPMJET interface to LEPTO--6.5.
       \end{quote}
  \item  dpm25nonu.f 
      \begin{quote}
        If DPMJET is to be applied only to hadronic and nuclear
	collisions, not to neutrino collisions, it is possible to drop
	the files leptonew.f, jetset74ku.f, dpm25lepto.f and
	dpm25qelpo.f and link the code instead to dpm25nonu.f, which
	contains only dummy routines.
       \end{quote}
\end{itemize}

\subsection{External input and output data files used } %
\begin{itemize}
  \item   GLAUBTAR.DAT, Unit 47 
     \begin{quote}
     Glauber model data, see input cards GLAUBERI, GLAUBERA.
       \end{quote}
  \item   NUCLEAR.BIN 
     \begin{quote}
     Binary file to read in nuclear data for use in the
     evaporation module.
       \end{quote}
  \item    qel.evt
     \begin{quote}
     Output file for the qel--events without the modifications
     of these events in DPMJET.
       \end{quote}
  \item    lepto.evt
     \begin{quote}
     Output file for the lepto--events without the modifications
     of these events in DPMJET.
       \end{quote}
\end{itemize}
%
\subsection{Sequence of subroutine calls for event generation }

\begin{enumerate}
  \item
     The generator subroutine KKINC
     monitors the sampling of a single event
     by one call of KKEVT, repeated FOZOKA calls and call(s) of the
     subroutine FICONF.
  \item
     The subroutine KKEVT performs the sampling
     of the primary projectile--target interaction.
     This includes
    \begin{itemize}
      \item
         sampling of the actual nucleon coordinates and assignment
	 of 'partners' in the $n$ elementary interactions
         between $n_p$ and $n_t$ nucleons from the projectile
         and the target, resp., done in SHMAKO~\cite{Shma88};
      \item
         sampling of Fermi momenta of all nucleons in colliding nuclei
         in FER4M;
      \item
         sampling of x-values for the appropriate quark systems
         (quarks, antiquarks and diquarks, resp.)
         from the interacting hadrons in the subroutine XKSAMP
         and assignment of quark flavors in FLKSAM;
      \item
         construction of color neutral parton-parton chains
         in the subroutines KKEVVV, KKEVSV, KKEVVS, KKEVSS and further
	 similar routines;
      \item
         hadronization of chains via the JETSET code
         monitored by the subroutine HADRKK;
      \item
        At energies below 3--5 GeV the code HADHAD is called as
	hadron--hadron event generator interfacing to the HADRIN
	code \cite{HADRIN}.
      \item
       Diffractive events are generated using a call to SDIFF. 
    \end{itemize}

  \item
     Alternatively the subroutines KKEVNU or KKEVLE treat the
     first step of the qel or lepto neutrino nucleus
     scattering.

         finally KKEVT, KKEVNU or KKEVLE  returns the control to KKINC.
  \item
     The intranuclear cascade is modeled by subsequent calls
     of FOZOKA for each secondary which did not leave the
     interacting nuclei. FOZOKA transports the particle either
     until the next interaction or until it leaves the nucleus.
  \item
     The  Routine FICONF is called to perform the evaporation
     step.
\end{enumerate}

%
%
\subsection{The test program DPMJET--II.5}

The event generator itself is supplied together with a test program
which may serve as an example for the application of the program.
By a call of the subroutine DMINIT 
(included in the source file dpm25nuc1.f)
the main program DPMJET--II.5 does the necessary 
initializations discussed above.
It also allows the definition/modification of model parameters via input
options explained in Appendix~A.
\\
The sample histogram routine DISTR (source file dpm25hist.f) generates
line printer output for average multiplicities, rapidity and
pseudorapidity distributions for created particles. 
The scoring procedure
demonstrates how the information on 
the event history stored in
the common /HKKEVT/ may be applied to extract the properties of the final
state particles.

%
%
\subsection{Remarks on the other codes applied in DPMJET--II.5:
   \hspace{20cm}  DIAGEN, DECAY, HADRIN 
    and RNDM}

All the external codes applied are well documented.
The generator DIAGEN \cite{Shma88} madified 
as described in \cite{Ranft99a}
samples configurations for
nucleus-nucleus interactions in the framework of the Glauber model,
i.e. spatial coordinates of nucleons in projectile and target nuclei,
resp., as well as the actual numbers of 'elementary' interactions
$n$, $n_p$ and $n_t$ (comp. sect.~2). DIAGEN routines are called
via SHMAKI (initialization) and SHMAKO (sampling of configurations),
both included in the file dpm25nuc7.f.
\\
A modified version of the resonance decay routine is defined by the
subroutine DECHKK treating the decay of resonances from the common
/HKKEVT/.
\\
The program HADRIN~\cite{HADRIN} for the description of hadron-nucleon
interactions below $5~GeV$ is well tested against data~\cite{HaeNSE1}.
In DPMJET--II.5 it is called via the subroutine FHAD (file dpm25hadri.f).
\\
The algorithm applied for the generation of uniform random numbers
(function RNDM) was described in ref. \cite{Marsaglia}.
The generated
sequence of pseudorandom numbers should be independent of the actual
hardware (if a real number has at least 24 significant bits in the
internal representation). This generator has a period of about
$2^{144}$. A few more comments and a test routine are provided in the
source file dpm25nulib.f.
%
%
\vspace{5mm}
\section*{Acknowledgements}
 I thank first of all my  collaborator Dr. H.--J. M\"ohring
 with whom together all previous versions of the code up to
 DTUNUC--1.03 were developped. 
The author  thanks Y.~Shmakov for supplying the DIAGEN code prior
to publication. Furthermore,  the support
of  CERN , the Department of Theoretical Physics in Lund, INFN,
Sezione
di Milano ,
INFN,LNF Frascati, LAPP Annecy,
The University Santiago de Compostelar,  INFN, Lab. Naz. del
Gran Sasso and the University of Siegen
where parts of the program were developped is acknowledged. The
code in the versions  described here was finalized  
at Siegen. 
The author acknowledges   
the fruitful collaborations with P.Aurenche, G.Battistoni, 
F.Bopp, M.Braun, A.~Capella,
R.Engel, A.Ferrari, C.Forti, K.H\"an\ss gen, K.Hahn, 
I.Kawrakov, C.Merino, N.Mokhov, H.J.M\"ohring, C.Pajares,
D.Pertermann, S.Ritter, S.Roesler, P.Sala   and
J.Tran~Thanh~Van on the Dual Parton Model in general. 

%
%
 \newpage
%
\bibliographystyle{zpc}
\bibliography{dpm11}
 
 \clearpage

%
%
\newpage
\section*{Appendix A~: \\ The test program DPMJET--II.5 }
The test program demonstrates the application of the event generating
routine KKINC and the extraction of information on the produced secondaries
from the common block /HKKEVT/ (in the routine DISTR).
Furthermore, it allows a simple redefinition of some important model
parameters. It may also be used to prepare data files
containing reaction specific information needed for the application of the
Glauber formalism.
In the case of event generation few standard histograms are constructed
from sampled events.
\\
All program activities are monitored by input options.
Each input option is identified by a code word and either changes
the default values of variables and/or demands some action.
Tab.~1 summarizes the standard input options.
All variables read from the input file have
default values.
\begin{table}[h]
%
\caption{ The most important options read by the subroutine DMINIT}
\begin{center}  \begin{tabular}{|l|l|}
%
\hline
Option \str  &    Description                                \\
	    &                                                \\
\hline
  TITLE \hspace{3cm}
	    & next card is the run title                     \\
  COMMENT   &   adds comments to the input data stream       \\
  PROJPAR   &   definition of the projectile                 \\
  TARPAR    &   definition of the target                     \\
  GLAUBERI  &   initialization of the Glauber formalism for   \\
	  &   hadron--nucleus interactions / data are written on unit 47\\
  GLAUBERA  &   initialization of the Glauber formalism for   \\
	  &   nucleus--nucleus interactions / data are written on unit 47\\
  XSECNUC   &   Calculation of  Glauber cross sections   \\
  HADRONIZ  &   selects  JETSET fragmentation of chains \\
  ENERGY    &   beam energy   ($GeV$; per nucleon for nuclei)   \\
  MOMENTUM  &   beam momentum ($GeV/c$; per nucleon for nuclei) \\
  CMENERGY  &   c.m. energy ($GeV$ per nucleon)   \\
  CENTRAL   &   central A-A collisions forced    \\
  TAUFOR    &   definition of the formation time parameter    \\
  SEADISTR  &   monitors the x-behaviour of the quark distributions \\
  SEAQUARK  &   monitors the x-behaviour of the sea--quark distributions \\
  SECINTER  &   demands secondary interactions \\
  SEASU3    &   defines $s^{sea}$ \\
  DIQUARKS  &   selects sea--diquarks at chain ends \\
  POPCORN   &   selects the popcorn mechanism \\
  POPCORCK  &   selects the Capella--Kopeliovich popcorn mechanism \\
  POPCORSE  &   selects the diquark breaking sea--quark mechanism \\
  CASADIQU  &   selects Casado diagram \\
  CRONINPT  &   selects the Cronin effect \\
  XCUTS     &   monitors cuts for x--sampling          \\
  FERMI     &   monitors assignment of Fermi momenta to nucleons  \\
  SINGDIFF  &   controls the inclusion of single diffractive events  \\
  SINGLECH  &   include Regge (single chain) contribution \\
  EVAPORAT  &   include nuclear evaporation  \\
  RANDOMIZ  &   reinitialize the RNDM (RM48) randon number generator  \\
  NOFINALE  &   Skip routine FICONF, which calculates the residual nuclei   \\
  START     &   start event sampling for h--h, h--A or A--A collisions\\
  NEUTRINO  &   start event sampling for qel neutrino collisions\\
  LEPTOEVT  &   start event sampling for lepto neutrino collisions\\
  STOP      &   stop the run                    \\
  PARTICLE  &   print table of available particles and properties \\
  POMTABLE  &   Write / Read a table for sampling multipomeron events\\
\hline
\end{tabular}   \end{center}
\end{table}

%
%
\newpage

\subsection*{A.1:  Description of input options and default
parameters}

{ \bf Table A--2}

{Particle codes, SDUM is the parameter to be given on
the PROJPAR input card. The internal codes and names were
originally introduced by the BAMJET \cite{BAMJET1,BAMJET} and DECAY
 \cite{DECAY} codes. These codes are also used by the 
 FLUKA \cite{FLUKA} and
DTUJET  \cite{DTUJET93,DTUJET93code} codes.In the output of
DPMJET, in COMMON HKKEVT, the particles are characterized with
the PDG code, the internal code is given in the EXTEVT common.
 (Not all particles used internally are given in
this Table.)
}
 \vskip 5mm
\begintable
SDUM |internal code | internal name | particle |  PDG number | charge |
q1 |q2 |q3    \cr   
  PROTON |   1 | P     |   $p^+               $  |   2212 |   1  |  2 |  2|  1      \cr   
  APROTON |   2 | AP    |   $ \bar{p}^-         $  |  -2212 |  -1  | -2 | -2| -1      \cr   
    |   3 | E-    |    $e^-              $  |     11 |  -1  |  0 | 0|  0      \cr   
    |   4 | E+    |  $ e^+               $  |    -11 |   1  |  0 | 0|  0      \cr   
    |   5 | NUE   |  $  \nu_e             $  |     12 |   0  |  0 | 0|  0      \cr   
    |   6 | ANUE  |  $  \bar{ \nu}_e       $  |    -12 |   0  |  0 | 0|  0      \cr   
    |   7 | GAM   |  $  \gamma            $  |     22 |   0  |  0 | 0|  0      \cr   
   NEUTRON |  8 | NEU   |  $ n^0               $  |   2112 |   0  |  2 | 1|  1      \cr   
   ANEUTRON |  9 | ANEU  |  $  \bar{n}0          $  |  -2112 |   0  | -2| -1| -1      \cr   
   |   10 | MUE+  |  $  \mu^+             $  |    -13 |   1  |  0 | 0|  0      \cr   
   |   11 | MUE-  |  $  \mu^-             $  |     13 |  -1  |  0 | 0|  0      \cr   
  KAONLONG |  12 | K0L   |  $ K_L^0             $  |    130 |   0  |  0 | 0|  0      \cr   
  PION+ |  13 | PI+   |  $  \pi^+             $  |    211 |   1  |  2| -1|  0      \cr   
  PION- |  14 | PI-   |  $  \pi^-             $  |   -211 |  -1  |  1| -2|  0      \cr   
  KAON+ |  15 | K+    |  $ K^+               $  |    321 |   1  |  2| -3|  0      \cr   
  KAON- |  16 | K-    |  $ K^-               $  |   -321 |  -1  |  3| -2|  0      \cr   
  LAMBDA |  17 | LAM   |  $  \Lambda^0         $  |   3122 |   0  |  0 | 0|  0      \cr   
  ALAMBDA |  18 | ALAM  |  $  \bar{ \Lambda}^0   $  |  -3122 |   0  |  0 | 0|  0      \cr   
  KAONSHRT |  19 | K0S   |  $ K_S^0             $  |    310 |   0  |  0 | 0|  0      \cr   
  SIGMA- |  20 | SIGM- |  $  \Sigma^-          $  |   3112 |  -1  |  1 | 1|  3      \endtable   
\newpage
\begintable
SDUM |internal code | internal name | particle |  PDG number | charge |
q1 |q2 |q3    \cr   
  SIGMA+ |  21 | SIGM+ |  $  \Sigma^+          $  |   3222 |   1  |  2 | 2|  3      \cr   
  SIGMAZER |  22 | SIGM0 |  $  \Sigma^0          $  |   3212 |   0  |  2 | 1|  3      \cr   
  PIZERO |  23 | PI0   |  $  \pi^0             $  |    111 |   0  |  2| -2|  0      \cr   
  KAONZERO |  24 | K0    |  $ K^0               $  |    311 |   0  |  1| -3|  0      \cr   
  AKAONZER |  25 | AK0   |  $  \bar{K}^0         $  |   -311 |   0  |  3| -1|  0      \cr   
  |     31 | ETA550 | $   \eta             $  |     221|    0 |  3| -3|  0      \cr   
  |     32 | RHO+77 | $   \rho^+           $  |     213 |   1 |  2| -1|  0      \cr   
  |     33 | RHO077 | $   \rho^0           $  |     113 |   0 |  2| -2|  0      \cr   
  |     34 | RHO-77 | $  \rho^-            $  |   -213  | -1  |  1| -2 | 0      \cr   
  |     35 | OM0783 | $  \omega            $  |    223  |  0  |  0|  0 | 0      \cr   
  |     36 | K*+892 | $ K^{*+}           $  |    323  |  1  |  2| -3|  0      \cr   
  |     37 | K*0892 | $ K^{*0}           $  |    313  |  0  |  1| -3|  0      \cr   
  |   38 | K*-892 | $ K^{*-}           $  |   -323 |  -1  |  3| -2|  0      \cr   
  |    39 | AK*089 | $  \bar{K^*}^0       $  |   -313 |   0  |  3| -1|  0      \cr   
  |    40 | KA+125 | $ K_1^+             $  |  10323 |   1  |  0|  0|  0      \cr   
  |    41 | KA0125 | $ K_1^0             $  |  10313 |   0  |  0|  0|  0      \cr   
  |    42 | KA-125 | $ K_1^-             $  | -10323 |  -1  |  0|  0|  0      \cr   
  |    43 | AKA012 | $  \bar{K_1}^0       $  | -10313 |   0  |  0|  0|  0      \cr   
  |    48 | S+1385 | $  \Sigma^{*+}       $  |   3224 |   1  |  0|  0|  0      \cr   
  |    49 | S01385 | $  \Sigma^{*0}       $  |   3214 |   0  |  0|  0|  0      \cr   
  |    50 | S-1385 | $  \Sigma^{*-}       $  |   3114 |  -1  |  0|  0|  0      \cr   
  |    53 | N*++12 | $  \Delta^{++}       $  |   2224 |   2  |  2|  2|  2      \cr   
  |    54 | N*+ 12 | $  \Delta^+          $  |   2214 |   1  |  2|  2|  1      \cr   
  |    55 | N*012  | $  \Delta^0          $  |   2114 |   0  |  2|  1 | 1      \cr   
  |    56 | N*-12  | $  \Delta^-          $  |   1114 |  -1  |  1|  1|  1      \cr   
  |    67 | AN--12 | $  \bar{ \Delta}^{--} $  |  -2224  | -2  | -2| -2 | 2      \cr   
  |    68 | AN*-12 | $  \bar{ \Delta}^-    $  |  -2214  | -1  | -2| -2| -1      \cr   
  |    69 | AN*012 | $  \bar{ \Delta}^0    $  |  -2114  |  0  | -2| -1| -1      \endtable   
\newpage
\begintable
SDUM |internal code | internal name | particle |  PDG number | charge |
q1 |q2 |q3    \cr   
  |    70 | AN*+12 | $  \bar{ \Delta}^+    $  |  -1114  |  1  | -1| -1 | 1      \cr   
  |    95 | ETA*   | $  \eta^ \prime       $  |    331  |  0  |  0|  0|  0      \cr   
  |    96 | PHI    | $  \phi              $  |    333  |  0  |  3| -3|  0      \cr   
  |    97 | TETA0  | $  \Xi^0             $  |   3322  |  0  |  2|  3|  3      \cr   
  |    98 | TETA-  | $  \Xi^-             $  |   3312  | -1  |  1|  3|  3      \cr   
  |    99 | ASIG-  | $  \bar{ \Sigma}^-    $  |  -3222  | -1  | -2| -2| -3      \cr   
  |   100 | ASIG0  | $  \bar{ \Sigma}^0    $  |  -3212  |  0  | -2| -1| -3      \cr   
  |   101 | ASIG+  | $  \bar{ \Sigma}^+    $  |  -3112  |  1  | -1| -1| -3      \cr   
  |   102 | ATETA0 | $  \bar{ \Xi}^0       $  |  -3322  |  0  | -2| -3 | 3      \cr   
  |   103 | ATETA+ | $  \bar{ \Xi}^+       $  |  -3312  |  1  | -1| -3| -3      \cr   
  |   104 | SIG*+  | $  \Sigma^{*+}       $  |   3224  |  1  |  2|  2|  3      \cr   
  |   105 | SIG*0  | $  \Sigma^{*0}       $  |   3214  |  0  |  2|  1|  3      \cr   
  |   106 | SIG*-  | $  \Sigma^{*-}       $  |   3114  | -1  |  1|  1|  3      \cr   
  |   107 | TETA*0 | $  \Xi^{*0}          $  |   3324  |  0  |  2|  3|  3      \cr   
  |   108 | TETA*  | $  \Xi^{*-}          $  |   3314  | -1  |  1|  3 | 3      \cr   
  |   109 | OMEGA- | $  \Omega^-          $  |   3334  | -1  |  3|  3 | 3      \cr   
  |   110 | ASIG*- | $  \bar{ \Sigma^*}^-  $  |  -3224  | -1  | -2| -2| -3      \cr   
  |   111 | ASIG*0 | $  \bar{ \Sigma^*}^0  $  |  -3214  |  0  | -2| -1| -3      \cr   
  |   112 | ASIG*+ | $  \bar{ \Sigma^*}+   $  |  -3114  |  1  | -1| -1| -3      \cr   
  |   113 | ATET*0 | $  \bar{ \Xi^*}^0     $  |  -3324  |  0  | -2| -3 | 3      \cr   
  |   114 | ATET*+ | $  \bar{ \Xi^*}^+     $  |  -3314  |  1  | -1| -3 | 3      \cr   
  |   115 | OMEGA+ | $  \bar{ \Omega}^+    $  |  -3334  |  1  | -3| -3 | 3      \cr   
  |   116 | D0     | $ D^0               $  |    421  |  0  |  4| -2|  0      \cr   
  |   117 | D+     | $ D^+               $  |    411  |  1  |  4| -1 | 0      \cr   
  |   118 | D-     | $ D^-               $  |   -411  | -1  |  1| -4|  0      \cr   
  |   119 | AD0    | $  \bar{D}^0         $  |   -421  |  0  |  2| -4|  0      \cr   
  |   120 | DS+    | $ D_s^+             $  |    431  |  1  |  4| -3|  0      \cr   
  |   121 | DS-    | $ D_s^-             $  |   -431  | -1  |  3| -4|  0      \endtable   
\newpage
\begintable
SDUM |internal code | internal name | particle |  PDG number | charge |
q1 |q2 |q3    \cr   
  |   122 | ETAC   | $  \eta_c            $  |    441  |  0  |  4| -4|  0      \cr   
  |   123 | D*0    | $ D^{*0}            $  |    423  |  0  |  4| -2|  0      \cr   
  |   124 | D*+    | $ D^{*+}            $  |    413  |  1  |  4| -1|  0      \cr   
  |   125 | D*-    | $ D^{*-}            $  |   -413  | -1  |  1| -4|  0      \cr   
  |   126 | AD*0   | $  \bar{D^*}^0       $  |   -423  |  0  |  2| -4|  0      \cr   
  |   127 | DS*+   | $ D^{*+}_s           $  |    433  |  1  |  4| -3 | 0      \cr   
  |   128 | DS*-   | $ D^{*-}_s           $  |   -433  | -1  |  3| -4|  0      \cr   
  |   129 | CHI1C  | $  \chi_1^c          $  |  20443  |  0  |  4| -4 | 0      \cr   
  |   130 | JPSI   | $ J/ \psi            $  |    443  |  0  |  0|  0|  0      \cr   
  |   131 | TAU+   | $  \tau^+            $  |    -15  |  1  |  0|  0|  0      \cr   
  |   132 | TAU-   | $  \tau^-            $  |     15  | -1  |  0|  0|  0      \cr   
  |   133 | NUET   | $  \nu_ \tau          $  |     16   | 0  |  0|  0|  0      \cr   
  |   134 | ANUET  | $  \bar{ \nu}_ \tau    $  |    -16  |  0  |  0|  0|  0      \cr   
  |   135 | NUEM   | $  \nu_ \mu           $  |     14  |  0  |  0|  0|  0      \cr   
  |   136 | ANUEM  | $  \bar{ \nu}_ \mu     $  |    -14  |  0  |  0|  0|  0      \cr   
  |   137 | LAMC+  | $  \Lambda_c^+       $  |   4122  |  1  |  2|  1|  4      \cr   
  |   138 | XIC+   | $  \Xi_c^+           $  |   4232  |  1  |  2|  3|  4      \cr   
  |   139 | XIC0   | $  \Xi_c^0           $  |   4132  |  0  |  1|  3|  4      \cr   
  |   140 | SIGC++  |$  \Sigma_c^{++}     $  |   4222  |  2  |  2|  2|  4      \cr   
  |   141 | SIGC+  | $  \Sigma_c^+        $  |   4212  |  1  |  0|  0|  0      \cr   
  |   142 | SIGC0  | $  \Sigma_c^0        $  |   4112  |  0  |  1|  1|  4      \cr   
  |   143 | S+     | $  \Xi^{ \prime +}_c    $  |   4322  |  1  |  0|  0|  0      \cr   
  |   144 | S0     | $  \Xi^{ \prime 0}_c    $  |   4312  |  0  |  0|  0|  0      \cr   
  |   145 | T0     | $  \Omega_c^0        $  |   4332  |  0  |  3|  3 | 4      \cr   
  |   146 | XU++   | $  \Xi_{cc}^{++}     $  |   4422  |  2  |  2|  4|  4      \cr   
  |   147 | XD+    | $  \Xi_{cc}^+        $  |   4412   | 1  |  1|  4|  4      \cr   
  |   148 | XS+    | $  \Omega_{cc}^+     $  |   4432  |  1  |  3|  4 | 4      \cr   
  |   149 | ALAMC- | $  \bar{ \Lambda}_c^- $  |  -4122   | 1  | -2| -1 | 4      \endtable   
\newpage
\begintable
SDUM |internal code | internal name | particle |  PDG number | charge |
q1 |q2 |q3    \cr   
  |   150 | AXIC-  | $ \bar{ \Xi}_c^-      $  |  -4232   | 1  | -2| -3| -4      \cr   
  |   151 | AXIC0  | $ \bar{ \Xi}_c^0      $  |  -4132  |  0  | -1| -3| -4      \cr   
  |   152 | ASIGC--| $ \bar{ \Sigma}_c^{--}$  |  -4222  | -2  | -2| -2| -4      \cr   
  |   153 | ASIGC- | $  \bar{ \Sigma}_c^-  $  |  -4212   | 1  |  0|  0|  0      \cr   
  |   154 | ASIGC0 | $  \bar{ \Sigma}_c^0  $  |  -4112   | 0  | -1| -1| -4      \cr   
  |   155 | AS-    |$ \bar{ \Xi}^{ \prime -}_c$  |  -4322   | 1  |  0|  0|  0      \cr   
  |   156 | AS0    |$ \bar{ \Xi}^{ \prime 0}_c$  |  -4312   | 0  |  0|  0|  0      \cr   
  |   157 | AT0    | $  \bar{ \Omega}_c^0  $  |  -4332   | 0  | -3| -3| -4      \cr   
  |   158 | AXU--  | $  \bar{ \Xi}_{cc}^{--}$ |  -4422  | -2  | -2| -4| -4      \cr   
  |   159 | AXD-   | $ \bar{ \Xi}_{cc}^-   $  |  -4412   | 1  | -1| -4| -4      \cr   
  |   160 | AXS    | $  \bar{ \Omega}_{cc}^-$ |  -4432  | -1  | -3| -4| -4      \cr   
  |   161 | C1*++  | $  \Sigma^{*++}_c    $ |   4224  |  2  |  2|  2|  4      \cr   
  |   162 | C1*+    |$  \Sigma^{*+}_c       $ |   4214  |  1  |  2|  1|  4      \cr   
  |   163 | C1*0   | $  \Sigma^{*0}_c       $ |   4114  |  0  |  1|  1|  4      \cr   
  |   164 | S*+    | $  \Xi^{*+}_c          $ |   4324  |  1  |  2|  3|  4      \cr   
  |   165 | S*0    | $  \Xi^{*0}_c          $ |   4314  |  0  |  1|  3|  4      \cr   
  |   166 | T*0    | $  \Omega^{*0}_c       $ |   4334  |  0  |  3|  3|  4      \cr   
  |   167 | XU*++  | $  \Xi^{*++}_{cc}      $ |   4424  |  2  |  2|  4|  4      \cr   
  |   168 | XD*+   | $  \Xi^{*+}_{cc}       $ |   4414  |  1  |  1|  4|  4      \cr   
  |   169 | XS*+   | $  \Omega^{*+}_{cc}    $ |   4434  |  1  |  3|  4|  4      \cr   
  |   170 | TETA++ | $  \Omega^{*++}_{ccc}  $ |   4444   | 2  |  4|  4 | 4      \cr   
  |   171 | AC1*-- | $ \bar{ \Sigma}^{*--}_c $ |  -4224   | 2  | -2| -2| -4      \cr   
  |   172 | AC1*-  | $  \bar{ \Sigma}^{*-}_c $ |  -4214  | -1  | -2| -1| -4      \cr   
  |   173 | AC1*0  | $  \bar{ \Sigma}^{*0}_c $ |  -4114  |  0  | -1| -1| -4      \cr   
  |   174 | AS*-   | $  \bar{ \Xi}^{*-}_c    $ |  -4324  | -1  | -2| -3| -4      \cr   
  |   175 | AS*0   | $  \bar{ \Xi}^{*0}_c    $ |  -4314  |  0  | -1| -3| -4      \cr   
  |   176 | AT*0   | $  \bar{ \Omega}^{*0}_c $ |  -4334  |  0  | -3| -3| -4      \cr   
  |   177 | AXU*-- | $  \bar{ \Xi}^{*--}_{cc}$ |  -4424  | -2  | -2| -4| -4      \cr   
  |   178 | AXD*-  | $  \bar{ \Xi}^{*-}_{cc} $ |  -4414   | 1  | -1| -4| -4      \cr   
  |   179 | AXS*-  | $  \bar{ \Omega}^{*-}_{cc}$|  -4434  | -1  | -3| -4| -4      \cr   
  |   180 | ATET-- | $ \bar{ \Omega}^{*--}_{ccc}$ |  -4444|  -2 | -4| -4| -4      \endtable

\newpage
All input records of DPMJET--II.5 have the following form:
\begin{verbatim}
	      CODEWD, (WHAT(I), I = 1,6), SDUM
	      FORMAT (A8, 2X, 6E 10.0, A8)
\end{verbatim}
\noindent
In the following we describe the meaning of the corresponding
variables for the standard input options
in the same order as the code words CODEWD are listed in Tab.~A--1.

Please note: further input options are described in the
Appendices C and D.

We give also the default values of the parameters .
\begin{itemize}
   \item   Code word = 'TITLE'
     \begin{quote}
	This option card must be followed by a card giving a run title,
	which will be reproduced in the output.
     \end{quote}
   \item   Code word = 'COMMENT'
     \begin{quote}
	This option allows to add comments to the input file
		       at arbitrary positions.
     \begin{description}
      \item[WHAT(1):]  number of comment cards following this card. \\
		       default : 1.0
     \end{description}
     \end{quote}
   \item   Code word = 'PROJPAR'
     \begin{quote}
	 This card defines the type of the projectile;      \\
	 if given it has to be included \underline{before} the
	 MOMENTUM/ENERGY option(s).
     \begin{description}
       \item[SDUM:] defines the projectile to be a hadron if given;
		    for naming conventions see Table 2.  \\
		    If SDUM is given WHAT(1) and WHAT(2)
		      need no specification;
		    for projectile nuclei SDUM has no meaning.
       \item[WHAT(1):]  mass number of projectile nucleus - IP
       \item[WHAT(2):]  atomic number of projectile nucleus - IPZ
     \end{description}
     default: incident proton (IP=1, IPZ=1).
    \end{quote}
   \item   Code word = 'TARPAR'
     \begin{quote}
       This card defines the type of the target nucleus.
     \begin{description}
       \item[WHAT(1):]  mass number of projectile nucleus - IT
       \item[WHAT(2):]  atomic number of projectile nucleus - ITZ
     \end{description}
      default: Nitrogen $N$ target ( IT=14, ITZ=7 )
     \end{quote}
%
   \item   Code word = 'GLAUBERI'
     \begin{quote}
       Requests the initialization of the Glauber formalism
	 for \underline{hadron}--nucleus interactions;  \\
	 the target nucleus has to be defined by the code word TARPAR
	 in advance;                                               \\
	 tables of impact parameter distributions for b--sampling are
	 written to unit 47 for several momenta
	 ($p_{lab}=\sqrt{10}^{(i+1)}; i=1,...,24$)
	   and different projectiles ($p, \pi^+$);           \\
	 WHAT(1) = JGLAUB
		   JGLAUB = 1 Calculation GLAUBTAR.DAT file (default)
     \end{quote}
%
%
   \item   Code word = 'GLAUBERA'
     \begin{quote}
       Requests the initialization of the Glauber formalism
	 for \underline{nucleus}--nucleus interactions;  \\
	 the target nucleus has to be defined by the code word TARPAR
	 in advance;                                               \\
	 the projectile nucleus has to be defined by the code word PROJPAR
	 in advance;                                               \\
	 tables of impact parameter distributions for b--sampling are
	 written to unit 47 for several momenta
	 ($p_{lab}=\sqrt{10}^{(i+1)}; i=1,...,24$)
	   and different projectiles ;           \\
	 WHAT(1) = JGLAUB
		   JGLAUB = 1 Calculation GLAUBTAR.DAT file (default)
     \end{quote}
%
%
   \item   Code word = 'XSECNUC'
     \begin{quote}
       Requests the calculation of Glaubercross sections 
	 for hadron (nucleus) --nucleus interactions;  \\
	 the target nucleus has to be defined by the code word TARPAR
	 in advance;                                               \\
	 the projectile nucleus has to be defined by the code word PROJPAR
	 in advance;                                               \\
	 WHAT(1) = ECMUU lowest CMS energy\\
	 WHAT(2) = ECMOO highest CMS energy\\
	 WHAT(3) = NGRITT number of CMS energy points\\
	 WHAT(4) = NEVTT number of Monte Carlo events for each cross
	 section calculation\\
     \end{quote}
%
%
   \item   Code word = 'HADRONIZ'
     \begin{quote}
         Selects  JETSET fragmentation of soft chains
     \begin{description}
      \item[WHAT(1):]  IHADRZ  (default: 2)   \\
                      IHADRZ = 2 selects JETSET fragmentation.  \\
                      IHADRZ = 11 selects an alternative JETSET fragmentation.
     \end{description}
     \end{quote}
   \item   Code word = 'ENERGY'
     \begin{quote}
	 This card defines the energy of the projectile
	 in the target rest system.
	 For incident nuclei the energy per nucleon is expected. \\
	 NOTE: only one of the ENERGY and the MOMENTUM options
	 is necessary, the last defined option is applied;
	 both these options are to be given
	 \underline{after} the PROJPAR definition.
     \begin{description}
      \item[WHAT(1):]  projectile energy in $GeV$   \\
     \end{description}
     \end{quote}
   \item   Code word = 'MOMENTUM'
     \begin{quote}
       This card defines the momentum of the projectile
       in the target rest system.
       For incident nuclei the momentum per nucleon is expected.\\
	 NOTE: only one of the ENERGY and the MOMENTUM options
	 is necessary, the last defined option is applied;
	 both these options are to be given
	 \underline{after} the PROJPAR definition.
     \begin{description}
      \item[WHAT(1):]   projectile momentum in $GeV/c$;
      \item[Default:] 100 000 GeV/c
      \end{description}
     \end{quote}
   \item   Code word = 'CMENERGY'
     \begin{quote}
	 Same as for code word 'ENERGY', but WHAT(1) defines the energy in
	 the hadron/nucleon--nucleon c.m. system.
     \end{quote}
   \item   Code word = 'CENTRAL'
     \begin{quote}
	This code word forces central nucleus-nucleus collisions,
	i.e. most nucleons of the projectile nucleus are forced to
	interact. The actual requirement depends on the atomic number
	of both the projectile and the target nuclei and is defined
	in the subroutine KKEVT (source file dpm25nuc2.f,
	after CALL SHMAKO ).
	Furthermore, the actual impact parameter is set near 
	to zero for this
	case in subroutine MODB (source file dpm25nuc7.f).
      \begin{description}
       \item[WHAT(1):]= ICENTR default: $0.0$, i.e. no forcing.\\
               ICENTR=0 : normal collisions\\
	       ICENTR=1 : central collisions with impact parameter
	       condition in dpm25nuc7.f and NA condition in
	       dpm25nuc2.f\\
	       ICENTR=2 : only  NA condition in dpm25nuc2.f\\
	       ICENTR=3 : less central condition  for Pb--Pb\\
	       ICENTR=10: peripheral collisions
      \end{description}
      Please note: The definition of central collisions is not
      unique, see the actual definitions in routine KKEVT
      (dpm25nuc2.f and dpm25nuc7.f). 
     \end{quote}
%
%
   \item   Code word = 'TAUFOR'
     \begin{quote}
	This option defines the formation time parameter controlling
	the intranuclear cascade.
	Additionally it allows to restrict the number
	of generations of secondary cascade interactions.
      \begin{description}
       \item[WHAT(1):]   formation time in $fm/c$;\\
			 default: $105~fm/c$
       \item[WHAT(2):]   maximum number of allowed generations
			 of secondary interactions;~~~~default: 0
       \item[WHAT(3):]   monitors the definition of the formation time
		       actually applied (comp. Subsects. 2.3 and 3.3.3):\\
		      WHAT(3)=1 : $p_{\perp}$--dependent formation time , (default)\\
		      WHAT(3)=2 : constant formation time
      \end{description}
      Please note, for h--h interactions, one should use WHAT(2)
= 0. ! The default corresponds to a supression of the formation
zone cascade. To apply the formation zone cascade, the
recommended values are (TAUFOR   5.   25.   1.).
     \end{quote}
%
   \item   Code word = 'SEADISTR'
     \begin{quote}
	 This option card defines properties
	 of the quark distributions,
	 which are of the general form~~~~
	   $ q(x) \propto x^{-\alpha} (1-x)^{\beta} $.
      \begin{description}
      \item[WHAT(1):] no meaning, $\alpha^{sea}$ is now
      controlled by the SEAQUARK card. (default $\alpha^{sea}$ =
      0.5)
		       default : $\alpha^{sea}=1.0$
      \item[WHAT(2):]  $\beta^{nuc}$ for valence-quark distributions
			     of nucleons;                  \\
		       default : $\beta^{nuc}=3.5$
      \item[WHAT(3):]  $\beta^{mes}$ for valence-quark distributions
			     of mesons;                  \\
		       default : $\beta^{mes}=1.11$
      \item[WHAT(4):]  no meaning; \\
      \end{description}
      NOTE: for reasons of the sampling efficiency the parameters for the
      sea distribution are fixed to $\alpha^{sea}$=either 1.0 or 0.5  
      and
      $\beta^{sea}=0.0$ in the present version of the subroutine XKSAMP
      and cannot be changed easily by the user.
     \end{quote}
%
%
   \item   Code word = 'SEAQUARK'
     \begin{quote}
	 This option card defines properties
	 of the quark distributions,
	 which are of the general form~~~~
	   $ q(x) \propto x^{-\alpha} (1-x)^{\beta} $.
      \begin{description}
      \item[WHAT(1):]  $\alpha^{sea}$ for sea-quark distributions
 		     possible values: 0.5 and 1.;   \\
		       default : $\alpha^{sea}=0.5$
      \end{description}
      NOTE: for reasons of the sampling efficiency the parameters for the
      sea distribution are fixed to $\alpha^{sea}$=either 1.0 or 0.5  
      and
      $\beta^{sea}=0.0$ in the present version of the subroutine XKSAMP
      and cannot be changed easily by the user.
     \end{quote}
%
%
   \item   Code word = 'SECINTER'
     \begin{quote}
        This option demands secondary interactions, the use of
	this option is at present only recommended for heavy ion
	collisions at SPS energies.
      \begin{description}
       \item[WHAT(1):] = 1. secondary interactions demanded  \\
			 default: 0
      \end{description}
      SECINTER should only be used for heavy ion collisions
      at CERN--SPS energies.
     \end{quote}
   \item   Code word = 'SEASU3'
     \begin{quote}
        This option determines the strange quark fraction
$s^{sea}$ at the sea--quark chain ends.
      \begin{description}
       \item[WHAT(1):]   $s^{sea}$,  \\
			 default: 0.5
      \end{description}
     \end{quote}
   \item   Code word = 'DIQUARKS'
     \begin{quote}
     This option determines diquarks at sea chain ends
      \begin{description}
       \item[WHAT(1):]   IDIQUA,  \\
			 default: 1
		 IDIQUA=0 no diquarks at Glauber sea chain ends\\
		 IDIQUA=1  diquarks at Glauber sea chain ends\\
       \item[WHAT(2):]   IDIQUU,  \\
			 default: 1
		 IDIQUU=0 no diquarks at unitary sea chain ends\\
		 IDIQUU=1  diquarks at unitary sea chain ends\\
       \item[WHAT(3):]   AMEDD,  \\
			 default: 0.9
		 1.-AMEDD is the fraction of sea chain ends with a
		 sea--diquark\\
      \end{description}
     \end{quote}
   \item   Code word = 'POPCORN'
     \begin{quote}
        This option determines the  popcorn mechanism for 
 JETSET fragmentation.
      \begin{description}
       \item[WHAT(1):]   PDB,  \\
			 default: 0.10 \\
                JETSET: PDB gives the fraction of diquarks
fragmenting directly into baryons. \\
                PDB = 0 switches off the popcorn
mechanism.
      \end{description}
     NOTE: POPCORN should appear before the HADRONIZE card.
     \end{quote}
   \item   Code word = 'POPCORCK'
     \begin{quote}
        This option determines the CK (Capella--Kopeliovich)
	popcorn mechanism for 
 JETSET fragmentation.
      \begin{description}
       \item[WHAT(1):]   IJPOCK,  \\
			 default: 0 \\
                IJPOCK=0 switches off the CK popcorn mechanism \\
mechanism.
       \item[WHAT(2):]   PDBCK,  \\
			 default: 0.00 \\
                 PDBCK gives the fraction of diquarks
 with CK mechanism \\
      \end{description}
     NOTE: POPCORCK should appear before the HADRONIZE card.
     \end{quote}
   \item   Code word = 'POPCORSE'
     \begin{quote}
        This option determines the CK (Capella--Kopeliovich)
	popcorn mechanism for 
 JETSET fragmentation.
      \begin{description}
       \item[WHAT(2):]   PDBSE,  \\
			 default: 0.45 \\
                 PDBSE gives the probability for a  diquark to be 
split by Glauber sea quarks \\
       \item[WHAT(2):]   PDBSEU,  \\
			 default: 0.45 \\
                 PDBSEU gives the probability for a  diquark to be 
split by unitary sea quarks \\
      \end{description}
     NOTE: POPCORSE should appear before the HADRONIZE card.
     \end{quote}
   \item   Code word = 'CASADIQU'
     \begin{quote}
        This option determines the  CASADO diagram
      \begin{description}
       \item[WHAT(1):]   ICASAD,  \\
			 default: 1 \\
                ICASAD=0 switches off the Casado mechanism \\
                ICASAD=1 switches on the Casado mechanism \\
       \item[WHAT(2):]   CASAXX,  \\
			 default: 0.50 \\
                CASAXX gives the probability for the Casado mechanism\\ 
      \end{description}
     \end{quote}
   \item   Code word = 'CRONINPT'
     \begin{quote}
        This option determines the  Cronin effect.
      \begin{description}
       \item[WHAT(1):]   MKCRON,  \\
			 default: 1. \\
          MKCRON = 0 switches off the Cronin effect.
       \item[WHAT(2):]   CRONCO,  \\
			 default: 0.64 \\
        CRONCO is the parameter in the parton multiple
scattering formula. 
      \end{description}
     \end{quote}
   \item   Code word = 'XCUTS'
     \begin{quote}
	 This option redefines the lower cuts for the sampling
	 of x-values to ensure minimum chain masses for hadronization
	 (used in XKSAMP).
      \begin{description}
      \item[WHAT(1)] = CVQ ;~~~$(x_q^{val})^{min} =
					 {\rm CVQ}/\sqrt{s}$,\\
		      lower cut for valence quarks;~~~~
		      default : ${\rm CVQ} = 1.8$
      \item[WHAT(2)] = CDQ ;~~~$(x_{qq}^{val})^{min} = {\rm CDQ}/
				    \sqrt{s}$,\\
		      lower cut for valence diquarks;~~~~
		      default : ${\rm CDQ} = 2.0$
      \item[WHAT(3)] = CSEA;~~~$(x_{q/\bar{q}}^{sea})^{min}
				       ={\rm CSEA}/\sqrt{s}$,\\
		      lower cut for sea quarks;~~~~
		      default : ${\rm CSEA} = 0.5$
      \item[WHAT(4)] = SSMIMA;~~~ $(x_{q/\bar{q}}^{sea})^{target}\cdot
			       (x_{q/\bar{q}}^{sea})^{project.}\cdot s
				\geq  (SSMIMA)^2$,\\
		      lower cut for the mass of sea-sea chains
		      applied in XKSAMP;   \\
		      default~: ${\rm SSMIMA} = 1.2~GeV.$
      \end{description}
     \end{quote}
%
   \item   Code word = 'FERMI'
     \begin{quote}
     \begin{description}
      \item[WHAT(1):]
	 Inclusion of Fermi momenta for nucleons if WHAT(1)$= 1.0$
		      (default)
      \item[WHAT(2):] FERMOD - scale factor for Fermi momentum
		      as calculated from zero temperature
		      Fermi distribution of nucleons; \\
		      default: FERMOD $= 0.6$.
      \item[WHAT(3):]
	 use zero temperature Fermi momentum distribution if
	 WHAT(3)$= 1.0$ (default); use distribution according to
	 Ref.\cite{Atti96} for WHAT(3)=2.
      \end{description}
     \end{quote}

   \item   Code word = 'SINGDIFF'
     \begin{quote}
	 This option controls the generation of single diffractive events.
     \begin{description}
      \item[WHAT(1):]   ISINGD;~~~~\\
                         ISINGD=1: Single diffraction included,\\
                         ISINGD=0: Single diffraction supressed,\\
                         ISINGD=2: Only single diffraction,\\
                         ISINGD=3: Only single diffraction, target excited,\\
                         ISINGD=4: Only single diffraction, projectile excited.
                        \\default: 1
      \end{description}
     \end{quote}
   \item   Code word = 'SINGLECH'
     \begin{quote}
     \begin{description}
      \item[WHAT(1):]   ISICHA;~~~~\\
                         ISINGD=1: include Regge (single chain) contributions,\\
                         ISINGD=0: single chains supressed,\\
                         ISINGD=2: Only single contribution,
                        \\default: 0
      \end{description}
      Please note: The single chain (Regge) contributions are
      only essential at low energies. they are at present only
      implemented for antibaryon and meson projectiles.
     \end{quote}
   \item   Code word = 'EVAPORAT'
     \begin{quote}
                                                                        
      Evaporation is performed if the EVAPORAT card is present
                                      Default:IEVAP=0 No evaporation
     \begin{description}
                                       
              \item [ what (1)] = IEVAP \\    
                   IEVAP  =  1:     evaporation is performed   \\                      
                   IEVAP    =  0: no evaporation   \\      
                                                                        
      \end{description}
      Actually the EVAPORAT card allows for more options, which normally
      are not needed by the user. See thee description in
      dpm25nuc1.f.
     \end{quote}
   \item   Code word = 'RANDOMIZ'
     \begin{quote}
                                                                        
     Re--initialization of RNDM (RM48) random number generator 
                                              Default:standard initialization

     ISEED1 and ISEED2 are printed at the end of each DPMJET run. They
     can also be generated by a RD2OUT call
     (AUAUAU=RD2OUT(ISEED1,ISEED2))					      
     \begin{description}
                                       
              \item [ what (1)] = ISEED1 \\    
            \item[ what (2)] = ISEED2    \\     
                                                                        
      \end{description}
     \end{quote}
   \item   Code word = 'NOFINALE'
     \begin{quote}
	This option skips the call of the routine FICONF, for
instance in heavy ion collisions;
	\begin{description}
	   \item[WHAT(I)  = 1.]  : no FICONF call, Default: 0.:
	\end{description}
     \end{quote}
%
 
   \item   Code word = 'START'
     \begin{quote}
	 This option starts the generation of 
	 events in h--h, h--A and A--A collisions
	 including the output of standard histograms.
     \begin{description}
      \item[WHAT(1):]   number of events to be sampled;~~~~default: 100
      \item[WHAT(2):]   the Glauber initialization is forced to be
			calculated
			in SHMAKI,\\
			i.e. no data read from file GLAUBTAR.DAT,
			if WHAT(2) = 1.0
      \end{description}
     \end{quote}
%
 
   \item   Code word = 'NEUTRINO'
     \begin{quote}
	 This option starts the generation of events
	 for quasi elastic (qel) neutrino--nucleus collisions
	 including the output of standard histograms.
     \begin{description}
      \item[WHAT(1):]   number of events to be sampled;~~~~default: 100
      \item[WHAT(2):]   neutrino type as defined in qel
 (1=$\nu_e$, 2=$\bar\nu_e$, 3=$\nu_{\mu}$, 4=$\bar\nu_{\mu}$,
 5=$\nu_{\tau}$, 6=$\bar\nu_{\tau}$)
      \end{description}
     \end{quote}
%
 
   \item   Code word = 'LEPTOEVT' 
     \begin{quote}
	 This option starts the generation of events
	 for  deep inelastic (lepto) neutrino--nucleus collisions
	 including the output of standard histograms.
     \begin{description}
      \item[WHAT(1):]   number of events to be sampled;~~~~default: 100
      \item[WHAT(2):]   neutrino type as defined in lepto
 (12=$\nu_e$, -12=$\bar\nu_e$, 14=$\nu_{\mu}$, -14=$\bar\nu_{\mu}$)
      \end{description}
     \end{quote}
   \item   Code word = 'STOP'
     \begin{quote}
	 This option stops the execution of the program.
     \end{quote}
   \item   Code word = 'PARTICLE'
     \begin{quote}
	 This card triggers a printout of all the particles defined
	 in the BAMJET-DECAY chain fragmentation,
	 including name conventions, quantum numbers and decay channels.
	 This are the internal particle codes used in DPMJET.
     \end{quote}
   \item   Code word = 'POMTABLE'
     \begin{quote}
     Only if the files dpm25pom.f and dpm25lap.f are linked
     instead of the files dtu25pom.f and dtu25lap.f.

     \begin{description}
      \item[WHAT(1):]  0 ( default) the file pomtab.dat is
      written. :1 The file pomtab.dat is read.
      \end{description}

     \end{quote}
\end{itemize}
%
%
\newpage
\subsection*{A.2:  Sample Input }

In the following several typical examples of input data are reproduced.
We give only the input cards which change the default parameters.

One of the  examples demonstrates the use of the 
test program to generate the
input data for the Glauber formalism (file GLAUBTAR.DAT) 
in $A$--$N$ collisions.

\subsubsection{Hadron-Hadron collisions}
Note: 
SINGDIFF includes diffractive events. XCUTS slightly changes
some defaults.
\begin{verbatim}

TITLE
 DPMJET p-p  200 GeV LAB  w. Diffr.
PROJPAR                                                               PROTON
TARPAR         1.0        1.
MOMENTUM  200.
SINGDIFF        1.       
XCUTS          0.70      2.0      0.30     1.201
START      1000000.       1.
STOP

\end{verbatim}
\subsubsection*{Hadron-Nucleus collisions}
Note: Here the NOFINALE and EVAPORATE options assure evaporation
and the working out of the residual nucleus. If this is
required, then also the TAUFOR card is needed in a form as given
to perform the FZIC. In hadron--nucleus collisions diffractive
events might be demanded like in the example or might be
supressed.

\begin{verbatim}

TITLE
 DPMJET p-N  100000 GeV CMS  w. Diffr.
PROJPAR                                                               PROTON
TARPAR        14.0        7.
CMENERGY  100000. 
NOFINALE    0.
EVAPORATE   1.
TAUFOR          5.0      25.
SINGDIFF        1.       
START      50000.       1.
STOP

\end{verbatim}

\subsubsection*{ Minimum bias nucleus-nucleus collisions}

Note: Here again the NOFINALE and EVAPORATE options assure evaporation
and the working out of the residual nuclei. If this is
required, then also the TAUFOR card is needed in a form as given
to perform the FZIC. In nucleus--nucleus collisions diffractive
events might be demanded or might be
supressed like in the example. In very high energy
nucleus--nucleus collisions the code might run more stable if
the Cronin effect is suppressed like in the excample given.

\begin{verbatim}

TITLE
 DPMJET Fe-N  10  TeV CMS  without Diffr.
PROJPAR       56.        26.                                        
TARPAR        14.0        7.
CMENERGY  10000.
NOFINALE    0.
EVAPORATE   1.
TAUFOR          5.      25.
SINGDIFF        0.       
CRONINPT        0.      0.00
START      10000.       1.
STOP


\end{verbatim}
\subsubsection*{Central  Nucleus-Nucleus collisions}
Note: In central nucleus-nucleus collisions diffractive events
should always be excluded like
in the  examples given. In both examples the FZIC and
evaporation is demanded (TAUFOR, NOFINALE, EVAPORATE options)
It is also possible to run nucleus--nucleus collisions without
FZIC and evaporation.
The last parameter on the XCUTS card is convenient to tune somewhat the
secondary particle multiplicity in nucleus--nucleus collisions.

\begin{verbatim}


TITLE
 DPMJET central S-S  200GeV LAB with secondary interactions
PROJPAR                                                               PROTON
PROJPAR       32.0       16.
TARPAR        32.0       16.
MOMENTUM  200.
NOFINALE    0.
EVAPORATE   1.
TAUFOR          5.0      25.
SECINTER    1.
CENTRAL         2.
SINGDIFF        0.       
CRONINPT        0.      0.00
XCUTS          1.80      2.0      0.50     2.50
START       10000.       1.
STOP


TITLE
 DPMJET central Pb-Pb  158 GeV LAB with secondary interaction
PROJPAR      207.0       82.
TARPAR       207.0       82.
MOMENTUM  158.
NOFINALE    0.
EVAPORATE   1.
TAUFOR          5.0      25.
SECINTER    1.
CENTRAL         2.
SINGDIFF        0.      
CRONINPT        1.      0.64
XCUTS          1.00      2.0      0.50     2.50
START       1000.       0.
STOP


\end{verbatim}

\subsubsection*{Cross section calculation}

\begin{verbatim}

TITLE
DPMJET p-Ar XSEC
PROJPAR                                                               PROTON
TARPAR        40.0       18.
XSECNUC          2.  1000000.       50.    2000.
STOP


TITLE
DPMJET Fe-N XSEC
PROJPAR                                                               PROTON
PROJPAR       56.0       26.
TARPAR        14.0        7.
XSECNUC          2.  1000000.       50.    20000.
STOP


\end{verbatim}
\subsubsection*{Input data for Glauber formalism}

\begin{verbatim}

TITLE
DPMJET-II Prepare Glauber data for different nucleus-N collisions
PROJPAR        1.        1. 
TARPAR        14.0        7.
GLAUBERI
PROJPAR        4.        2. 
TARPAR        14.0        7.
GLAUBERA
PROJPAR       12.        6. 
TARPAR        14.0        7.
GLAUBERA
PROJPAR       14.        7. 
TARPAR        14.0        7.
GLAUBERA
PROJPAR       24.        12.
TARPAR        14.0        7.
GLAUBERA
PROJPAR       28.        14.
TARPAR        14.0        7.
GLAUBERA
PROJPAR       40.        20.
TARPAR        14.0        7.
GLAUBERA
PROJPAR       48.        22.
TARPAR        14.0        7.
GLAUBERA
PROJPAR       56.0       26.
TARPAR        14.0        7.
GLAUBERA
STOP

\end{verbatim}

\subsubsection*{ qel neutrino interactions in DPMJET--II.5}

Please note: DPMJET does not consider neutrinos as projectile
particles, therefore, we use PROJPAR = PROTON. This is of course
not used later on.

\begin{verbatim}

TITLE
 DPMJET-II.5 5 GeV Neutrino(nu-e)-Ar  quasi-elastic interaction
PROJPAR                                                               PROTON 
TARPAR          40.0       18.
MOMENTUM          5.
NOFINALE          0.
EVAPORATE         1.
TAUFOR           5.0      25.
NEUTRINO       1000.      1. 
STOP

\end{verbatim}

Please note: DPMJET does not consider neutrinos as projectile
particles, therefore, we use PROJPAR = PROTON. This is of course
not used later on.

\subsubsection*{Deep inelastic (lepto) neutrino-nucleus interaction
in DPMJET--II.5}

\begin{verbatim}

TITLE
 DPMJET Deep inelastic Neutrino(nu-e)-Ar using lepto
PROJPAR                                                               PROTON 
TARPAR          40.0       18.
MOMENTUM         50.
NOFINALE          0.
EVAPORATE         1.
TAUFOR           5.0      25.
LEPTOEVT       1000.      12. 
STOP

\end{verbatim}
%
%
\newpage
\section*{Appendix B~: \\ Event history and the common blocks /HKKEVT/
and /EXTEVT/}

\subsection*{B.1: Structure of the common blocks }
During the generation of individual events several entries are scored
in the common blocks /HKKEVT/ and /EXTEVT/
characterizing subsequent stages of
the sampling process. Scored entries are, for instance,
initial state nucleons, partons and parton chains, decaying resonances
as well as final state particles. These entries are characterized
by their type, 4-momenta and coordinates;
additional pointers define 'parents' and 'daughters' of the
actual entry (if any).
 The structure of the /HKKEVT/ common block closely follows
the suggestions of Ref.~\cite{ZLEP,ZLEP1}.
Within the code there are extensive comments
explaining the variables used in this common block.
Below the common blocks are reproduced together with these comments.
\\
Note that for interactions potentially resulting in high multiplicities of
secondaries (i.e. very high energies and/or 
heavy ion--ion collisions) it
may become necessary to increase the dimension NMXHKK, the maximum
number of entries for a given event.
\begin{verbatim}
      PARAMETER (NMXHKK=49998)
      COMMON /HKKEVT/ NHKK,NEVHKK,ISTHKK(NMXHKK),IDHKK(NMXHKK),
     &                JMOHKK(2,NMXHKK),JDAHKK(2,NMXHKK),
     &                PHKK(5,NMXHKK),VHKK(4,NMXHKK),WHKK(4,NMXHKK)
      COMMON /EXTEVT/ IDRES(NMXHKK),IDXRES(NMXHKK),NOBAM(NMXHKK),
     &                IDBAM(NMXHKK),IDCH(NMXHKK),NPOINT(10)
C
C  Based on the proposed standard COMMON block (Sjostrand Memo 17.3,89)
C
C NMXHKK: maximum numbers of entries (partons/particles) that can be
C    stored in the common block.
C
C NHKK: the actual number of entries stored in current event. These are
C    found in the first NHKK positions of the respective arrays below.
C    Index IHKK, 1 <= IHKK <= NHKK, is used below to denote a given
C    entry.
C
C ISTHKK(IHKK): status code for entry IHKK, with following meanings:
C    = 0 : null entry.
C    = 1 : an existing entry, which has not decayed or fragmented.
C        This is the main class of entries which represents the
C        "final state" given by the generator.
C    = 2 : an entry which has decayed or fragmented and therefore
C        is not appearing in the final state, but is retained for
C        event history information.
C    = 3 : a documentation line, defined separately from the event
C        history. (incoming reacting
C        particles, etc.)
C    = 4 - 10 : undefined, but reserved for future standards.
C    = 11 - 20 : at the disposal of each model builder for constructs
C        specific to his program, but equivalent to a null line in the
C        context of any other program. One example is the cone defining
C        vector of HERWIG, another cluster or event axes of the JETSET
C        analysis routines.
C    = 21 - : at the disposal of users, in particular for event tracking
C        in the detector.
C
C IDHKK(IHKK) : particle identity, according to the Particle Data Group
C    standard.
C
C JMOHKK(1,IHKK) : pointer to the position where the mother is stored.
C    The value is 0 for initial entries.
C
C JMOHKK(2,IHKK) : pointer to position of second mother. Normally only
C    one mother exist, in which case the value 0 is used. In cluster
C    fragmentation models, the two mothers would correspond to the q
C    and qbar which join to form a cluster. In string fragmentation,
C    the two mothers of a particle produced in the fragmentation would
C    be the two endpoints of the string (with the range in between
C    implied).
C
C JDAHKK(1,IHKK) : pointer to the position of the first daughter. If an
C    entry has not decayed, this is 0.
C
C JDAHKK(2,IHKK) : pointer to the position of the last daughter. If an
C    entry has not decayed, this is 0. It is assumed that the daughters
C    of a particle (or cluster or string) are stored sequentially, so
C    that the whole range JDAHKK(1,IHKK) - JDAHKK(2,IHKK) contains
C    daughters. Even in cases where only one daughter is defined (e.g.
C    K0 -> K0S) both values should be defined, to make for a uniform
C    approach in terms of loop constructions.
C
C PHKK(1,IHKK) : momentum in the x direction, in GeV/c.
C PHKK(2,IHKK) : momentum in the y direction, in GeV/c.
C PHKK(3,IHKK) : momentum in the z direction, in GeV/c.
C PHKK(4,IHKK) : energy, in GeV.
C PHKK(5,IHKK) : mass, in GeV/c**2. For spacelike partons, it is allowed
C    to use a negative mass, according to PHKK(5,IHKK) = -sqrt(-m**2).
C
C VHKK(1,IHKK) : production vertex x position, in mm.
C VHKK(2,IHKK) : production vertex y position, in mm.
C VHKK(3,IHKK) : production vertex z position, in mm.
C VHKK(4,IHKK) : production time, in mm/c (= 3.33*10**(-12) s).
C
C WHKK(I,IHKK) gives positions and times in projectile frame,
C    the chains are created on the positions of the projectile nucleons
C    in the projectile frame (target nucleons in target frame);
C    both positions are threfore not completely consistent.
C    The times in the projectile frame ,however, are obtained by
C    a Lorentz transformtion from the lab system.
C
C         Entries to the /EXTEVT/ COMMON block
C
C IDRES(IHKK) gives the Mass number A in case of nuclear
C     fragments or residual nuclei (IDHKK(IHKK)=80000)
C
C IDXRES(IHKK) gives the charge Z in case of nuclear
C     fragments or residual nuclei (IDHKK(IHKK)=80000)
C
C NOBAM(IHKK) = 1 for particles resulting from projectile fragmentation
C             = 2 for particles resulting from target fragmentation
C
C IDBAM(IHKK) gives the internal dpmjet particle code (BAMJET code)
\end{verbatim}

%
%
\newpage
\subsection*{B.2: Conventions for the scoring of the event history in
		  /HKKEVT/}

In the following we briefly characterize the subsequent entries
to /HKKEVT/ together with the most important conventions
for their classification.
\begin{itemize}
  \item
     projectile hadron/nucleons; \\
     for projectile nuclei Fermi momenta in the projectile rest frame
     and coordinates within the nucleus are stored in the arrays
     PHKK and VHKK, resp., by the subroutine KKEVT; \\
     interacting and non-interacting nucleons have the status
     ISTHKK=11~and~13, resp.;\\
     nucleons wounded by the formation zone intranuclear cascade
     get status 17.
  \item
     target nucleons; \\
     Fermi momenta in the target rest frame
     and coordinates within the nucleus are defined in PHKK and VHKK,
     resp. (KKEVT);  \\
     interacting and non-interacting nucleons have the status
     ISTHKK=12~and~14, resp.;\\
     nucleons wounded by the formation zone intranuclear cascade
     get status 18.
  \item
     valence quarks / diquarks from the interacting projectile
     hadron/nucleon(s)
     defined in the subroutine XKSAMP (total number IXPV);  \\
     PHKK(3,...)=PHKK(4,...) contains the actual momentum fraction,
     VHKK the position of the 'mother' hadron;   \\
     defined status ISTHKK=21;
  \item
     sea quarks from interacting projectile hadrons
     defined in XKSAMP (total number IXPS);  \\
     PHKK(3,...)=PHKK(4,...) contains the actual momentum fraction,
     VHKK the position of the 'mother' hadron;   \\
     defined status ISTHKK=31;
  \item
     valence quarks / diquarks from interacting target nucleons
     defined in XKSAMP (number IXTV);  \\
     PHKK(3,...)=PHKK(4,...) contains the actual momentum fraction,
     VHKK the position of the 'mother' hadron;   \\
     defined status ISTHKK=22;
  \item
     sea quarks from interacting target nucleons defined in XKSAMP
     (number IXTS);  \\
     PHKK(3,...)=PHKK(4,...) contains the actual momentum fraction,
     VHKK the position of the 'mother' hadron;   \\
     defined status ISTHKK=32;
  \item
     characteristics of the individual parton-parton chains (before
     hadronization) from subroutines KKEVVV, KKEVSV, KKEVVS and KKEVSS;
     for each chain there are three entries: \\
     (1) two entries for the quark systems forming the chain; \\
       \hspace{25mm}
	 PHKK gives their 4-momenta; the status of the corresponding
	 quark system is increased by 100 as compared to the
	 previous entry from the subroutine XKSAMP
	 (i.e. ISTHKK=121,122,131 or 132 ,resp.);  \\
     (2) one entry for the complete chain; \\
       \hspace{25mm}
	 PHKK gives the total 4-momentum,
	 the 'particle' type for chains is defined to be IDHKK=88888; \\
	 the actual status ISTHKK points to the chain generating
	 subroutine: \\
	 ~~~~ISTHKK=3 for chains from  KKEVVV (constructed from
		  valence quark systems), \\
	 ~~~~ISTHKK=4 for chains from subroutine KKEVSV
		  (sea-valence chains), \\
	 ~~~~ISTHKK=5 for chains from subroutine KKEVVS
		  (valence-sea chains), \\
	 ~~~~ISTHKK=6 for chains from subroutine KKEVSS
		  (sea-sea chains); \\
  \item
     hadrons from the hadronization of chains, entries from
     subroutines HADRSS, HADRVS, HADRSV, HADRVV;  \\
     assignment of values to all arrays of /HKKEVT/,
     status ISTHKK=1;
  \item
     hadrons from resonance decay in JETSET or in subroutine DECHKK \\
     (presently called after completion of the primary interaction
     of the projectile treated in KKEVT); \\
     the status of decaying hadrons is changed to ISTHKK=2,
     added decay products have ISTHKK=1;
  \item
     hadrons from intranuclear cascade interactions
     (monitored by FOZOKL):\\
     the status of interacting secondary is changed to ISTHKK=2;
     interacting target nucleons get ISTHKK=18,
     final state hadrons have the status ISTHKK=1.
     \\
     Particular cases:
     \begin{itemize}
       \item[(i)]
	  If a given secondary interaction is found to be forbidden
	  because of the Pauli principle the initial state particles
	  are stored in /HKKEVT/ with their original properties, but
	  the actual position; so they may participate in further
	  intranuclear interactions.
       \item[(ii)]
	  One (or two) nucleons from a secondary interaction cannot
	  escape from the nuclear potential, but the particular
	  collision is \underline{not} forbidden by Pauli's principle
	  (i.e. several nucleons knocked out of the nucleus already):
	  \\
	  Store the nucleon(s) with the actually generated momentum
	  at the collision site, assigning the status ISTHKK=15~(16)
	  for interactions in the target (projectile) nucleus.
	  Those nucleons are available as target (projectile)
	  nucleons in subsequent steps of the intranuclear cascade
	  development.
       \item[(iii)]
	  Negative particles with energies too low to escape from the
	  potential are forced to be absorbed within the nucleus
	  (comp. Section 2 and Appendix); absorbed $\pi^-, K^-$ and
	  $\bar{p}$ are characterized by the status ISTHKK=19.
     \end{itemize}
  \item
  Evaporation nucleons, nuclear fragments and residual target
  nuclei from the interface to the evaporation mudule.
     \begin{itemize}
       \item[(i)]
       The excited residual nuclei before the evaporation step
       get ISTHKK($i$)=1000 and  IDHKK($i$)=80000 ,
        the mass number A and charge Z of the nucleus are given
	by IDRES($i$) and IDXRES($i$).
       \item[(ii)]
       The evaporation protons and neutrons are stored with
       ISTHKK($i$)=-1.
       \item[(iii)]
       The nuclear fragments from evaporation  are stored with
       ISTHKK($i$)=-1 and IDHKK($i$)=80000,
        the mass number A and charge Z of the nucleus are given
	by IDRES($i$) and IDXRES($i$).
       \item[(iv)]
       The deexcitation photons are stored with
       ISTHKK($i$)=-1. 
       \item[(v)]
       The stable residual nuclei after the evaporation step 
        are stored with
       ISTHKK($i$)=1001 and IDHKK($i$)=80000,
        the mass number A and charge Z of the nucleus are given
	by IDRES($i$) and IDXRES($i$).
     \end{itemize}
\end{itemize}
The information from this common block allows a rather detailed
reconstruction of the actual event history, and is particularly
useful for consistency tests and debugging.
An example is discussed in Subsection~B.3.

%
%
\newpage
\subsection*{B.3: Sample event history from the common block /HKKEVT/ }

In this subsection we discuss a typical event history as stored
in the common block /HKKEVT/ according to the conventions
discussed in the preceding subsection~B.2.
The corresponding event was
sampled for the interaction of a $200~GeV$ proton
with a Sulfur nucleus.
The following quantities from the common are listed for each entry:
the number $i$ of the entry, its status ISTHKK($i$) and type IDHKK($i$),
the pointers JMOHKK(2,$i$) and JDAHKK(2,$i$) to the 'mother' and
'daughter' particles, resp., and the array PHKK(5,$i$),
for final state particles containing the 4-momentum and their mass.
The last 4 integers are IDRES($i$), IDXRES($i$), NOBAM($i$) and
IDBAM($i$).
\\
The first 33 entries are made in the subroutine KKEVT
for the incident proton and the 40 nucleons from the Sulfur nucleus.
According to our conventions interacting hadrons get the
status 11~(projectile) and 12~(target) - there are two target nucleons
participating in the primary interaction (entries 2 and 25, i.e two
 neutrons). 
 There are some target nucleons which got the
status ISTHKK($i$)=18, this are target nucleons participating in
the formation zone intranuclear cascade. 
\begin{verbatim}
   1  11  2212   0   0  42  45      0.00      0.00      0.00      0.94      0.94  0  0 0   1
   2  12  2112   0   0  49  52     -0.08     -0.12     -0.10      0.96      0.94  0  0 0   8
   3  14  2212   0   0 108   0      0.02      0.08     -0.14      0.95      0.94  0  0 0   1
   4  14  2112   0   0   0   0      0.01     -0.05      0.09      0.95      0.94  0  0 0   8
   5  14  2112   0   0   0   0     -0.07     -0.06     -0.14      0.96      0.94  0  0 0   8
   6  14  2212   0   0   0   0     -0.04      0.15     -0.06      0.95      0.94  0  0 0   1
   7  14  2212   0   0   0   0     -0.04      0.15      0.03      0.95      0.94  0  0 0   1
   8  14  2212   0   0   0   0     -0.08      0.13      0.08      0.96      0.94  0  0 0   1
   9  14  2212   0   0   0   0     -0.03      0.11      0.01      0.95      0.94  0  0 0   1
  10  14  2212   0   0   0   0      0.09      0.05      0.13      0.95      0.94  0  0 0   1
  11  14  2112   0   0   0   0      0.12      0.04     -0.04      0.95      0.94  0  0 0   8
  12  14  2112   0   0   0   0     -0.04      0.15      0.05      0.95      0.94  0  0 0   8
  13  14  2112   0   0   0   0     -0.08      0.00     -0.13      0.95      0.94  0  0 0   8
  14  18  2112   0   0  92  95     -0.09     -0.02     -0.02      0.94      0.94  0  0 0   8
  15  14  2212   0   0   0   0      0.13     -0.09      0.05      0.95      0.94  0  0 0   1
  16  14  2212   0   0   0   0      0.02     -0.09      0.12      0.95      0.94  0  0 0   1
  17  14  2112   0   0   0   0      0.05      0.03      0.02      0.94      0.94  0  0 0   8
  18  18  2212   0   0  96  97      0.08     -0.05     -0.03      0.94      0.94  0  0 0   1
  19  18  2112   0   0 102 103      0.10      0.08     -0.09      0.95      0.94  0  0 0   8
  20  14  2212   0   0   0   0     -0.06      0.11      0.14      0.96      0.94  0  0 0   1
  21  14  2112   0   0   0   0      0.08     -0.03     -0.08      0.95      0.94  0  0 0   8
  22  18  2112   0   0 104 105     -0.11      0.04      0.01      0.95      0.94  0  0 0   8
  23  14  2212   0   0   0   0     -0.12     -0.01      0.04      0.95      0.94  0  0 0   1
  24  14  2112   0   0   0   0     -0.16     -0.07      0.01      0.96      0.94  0  0 0   8
  25  12  2112   0   0  43  46      0.01     -0.05      0.14      0.95      0.94  0  0 0   8
  26  18  2112   0   0  90  91      0.00     -0.13     -0.13      0.96      0.94  0  0 0   8
  27  14  2112   0   0   0   0      0.08     -0.01      0.01      0.94      0.94  0  0 0   8
  28  18  2212   0   0 106 107     -0.10     -0.15      0.07      0.96      0.94  0  0 0   1
  29  14  2212   0   0   0   0      0.06      0.02      0.16      0.95      0.94  0  0 0   1
  30  14  2112   0   0   0   0      0.13     -0.03     -0.08      0.95      0.94  0  0 0   8
  31  18  2212   0   0  98 101      0.03     -0.03     -0.15      0.95      0.94  0  0 0   1
  32  14  2212   0   0   0   0      0.10      0.02      0.13      0.95      0.94  0  0 0   1
  33  14  2212   0   0   0   0      0.01     -0.15     -0.11      0.96      0.94  0  0 0   1
\end{verbatim}
The next entries come from the subroutines XKSAMP and FLKSAM (in module
DPMNUC3) assigning $x$ values and flavor to the quarks
participating in the
interaction: The projectile proton is split into a ${d}$ valence quark,
an valence diquark $({uu})$
           (both with status ISTHKK=21), and two quarks
$d$ and $\bar{d}$ from a colorless sea pair (status ISTHKK=31).
The interacting target nucleons are split in valence quark-diquark
systems: the neutron (entry 2) into $d$-$(ud)$, the neutron (entry 25)
into $d$-$(ud)$. The $x$ values given for each quark system
as PHKK(3,$i$)=PHKK(4,$i$) add up to one for each individual hadron.
\begin{verbatim}
  34  21     1   1   0  48  48      0.00      0.00      0.18      0.18      0.00  0  0 0 391
  35  21  2203   1   0  51  51      0.00      0.00      0.32      0.32      0.00  0  0 0 410
  36  31     1   1   0  42  42      0.00      0.00      0.01      0.01      0.00  0  0 0 391
  37  31    -1   1   0  45  45      0.00      0.00      0.50      0.50      0.00  0  0 0 392
  38  22     1   2   0  52  52      0.00      0.00      0.13      0.13      0.00  0  0 0 391
  39  22  2103   2   0  49  49      0.00      0.00      0.87      0.87      0.00  0  0 0 410
  40  22     1  25   0  46  46      0.00      0.00      0.11      0.11      0.00  0  0 0 391
  41  22  2103  25   0  43  43      0.00      0.00      0.89      0.89      0.00  0  0 0 410
\end{verbatim}
The following entries come from the routines constructing the
colorless parton-parton chains. First the chains are considered
(in subroutine KKEVSV) which are composed from projectile sea quarks
and target valence quarks. In this example there are two such chains:
$d$-$(ud)$ and $\bar{d}$-$u$. The entries are made subsequently
for the partons forming the chain and the constructed chain itself,
with the array PHKK(5,$i$) containing the corresponding 4-momenta
and the mass, resp. The 4-momenta are defined
in the cms system of the incident hadron and a single nucleon
(without Fermi momentum).
By definition these chains get the status ISTHKK=4, the status
of quarks/quark systems at chain ends is increased by 100
compared to their previous entries.

Please note: quarks, diquarks and chains have no BAMJET code,
therefore the last entry in the line (IDBAM($i$)) picks up values
without significance
\begin{verbatim}
  42 131     1  36   1  44  44     -0.06     -0.10      0.02      0.19      0.00  0  0 0 391
  43 122  2103  41  25  44  44      0.03     -0.03     -7.36      7.41      0.00  0  0 0 410
  44   4 88888  42  43  54  57     -0.03     -0.13     -7.34      7.59      1.94  0  0 0   0
  45 131    -1  37   1  47  47     -0.03      0.00     -0.05      0.96      0.00  0  0 0 392
  46 122     1  40  25  47  47      0.23     -0.07      3.90      4.84      0.00  0  0 0 391
  47   4 88888  45  46  58  67      0.20     -0.07      3.86      5.80      4.33  0  0 0   0
\end{verbatim}
In the same manner the next entries describe the chains constructed from
valence quark systems of both the projectile and a target nucleon
(neutron in this example) in the subroutine KKEVVV.
Entries for all chains are marked by the identity IDHKK=88888,
valence-valence chains get the status ISTHKK=3.
\begin{verbatim}
  48 121     1  34   1  50  50     -0.15     -0.02      1.54      1.72      0.00  0  0 0 391
  49 122  2103  39   2  50  50      0.17     -0.11     -9.25      9.43      0.00  0  0 0 410
  50   3 88888  48  49  68  81      0.03     -0.13     -7.71     11.15      8.05  0  0 0   0
  51 121  2203  35   1  53  53     -0.25     -0.01     -1.15      1.45      0.00  0  0 0 410
  52 122     1  38   2  53  53     -0.13     -0.26      2.75      3.07      0.00  0  0 0 391
  53   3 88888  51  52  82  89     -0.38     -0.27      1.59      4.53      4.21  0  0 0   0
\end{verbatim}
The next entries are generated by the routines HADRxx monitoring the
chain decay.

First the decay of the (two) sea-valence chains is
handled via HADRSV.
With ISTHKK($i$)=1 we find the final stable hadrons, with
ISTHKK($i$)=2 we find the resonances, which have decayed already
within the JETSET fragmentation code. The stable hadrons resulting from
resonance dacay point to the decaying resonance (IMOHKK(1,i)). 
Also stable hadrons might
have ISTHKK($i$)=2, in this case they have interacted
subsequently in the formation zone cascade.

After event completion the 4-momenta are given in the laboratory system.
\begin{verbatim}
  54   1  -211  44   0   0   0      0.25     -0.59      0.69      0.95      0.14  0  0 0  14
  55   2  2214  44   0   0   0     -0.28      0.46      1.55      2.02      1.17  0  0 0  54
  56   1  2212  55   0   0   0     -0.07      0.38      1.30      1.65      0.94  0  0 2   1
  57   1   111  55   0   0   0     -0.21      0.07      0.22      0.34      0.13  0  0 2  23
  58   2   223  47   0   0   0      0.20     -0.04      7.99      8.03      0.79  0  0 0  35
  59   1  -211  47   0   0   0      0.55     -0.20      5.85      5.88      0.14  0  0 2  14
  60   1   211  47   0   0   0      0.10     -0.15     12.38     12.38      0.14  0  0 2  13
  61   2   113  47   0   0   0     -1.00      0.41     47.65     47.67      0.69  0  0 0  33
  62   1   111  47   0   0   0      0.35     -0.09     25.78     25.78      0.13  0  0 2  23
  63   1   211  58   0   0   0      0.13      0.26      2.42      2.44      0.14  0  0 2  13
  64   1  -211  58   0   0   0      0.13     -0.08      2.36      2.37      0.14  0  0 2  14
  65   1   111  58   0   0   0     -0.06     -0.22      3.21      3.22      0.13  0  0 2  23
  66   1   211  61   0   0   0     -0.54     -0.12     20.43     20.44      0.14  0  0 2  13
  67   1  -211  61   0   0   0     -0.45      0.52     27.22     27.23      0.14  0  0 2  14
  68   2   221  50   0   0   0      0.06     -0.36     20.51     20.52      0.55  0  0 0  31
  69   2   223  50   0   0   0      0.48      0.02      6.32      6.38      0.78  0  0 0  35
  70   2   113  50   0   0   0      0.01      0.48      8.05      8.11      0.77  0  0 0  33
  71   1  -211  50   0   0   0     -0.64      0.44      0.79      1.11      0.14  0  0 0  14
  72   1  2212  50   0   0   0      0.11     -0.69      1.24      1.70      0.94  0  0 0   1
  73   1  -211  68   0   0   0      0.07     -0.12      9.17      9.17      0.14  0  0 2  14
  74   1   211  68   0   0   0      0.10     -0.15      4.20      4.20      0.14  0  0 2  13
  75   1   -11  68   0   0   0     -0.06     -0.08      2.68      2.68      0.00  0  0 0   4
  76   1    11  68   0   0   0     -0.05     -0.01      4.46      4.46      0.00  0  0 0   3
  77   2   211  69   0  90  91      0.25      0.28      3.43      3.46      0.14  0  0 0  13
  78   1  -211  69   0   0   0      0.16     -0.18      1.06      1.10      0.14  0  0 2  14
  79   1   111  69   0   0   0      0.07     -0.08      1.77      1.78      0.13  0  0 2  23
  80   1  -211  70   0   0   0     -0.10      0.51      4.90      4.93      0.14  0  0 0  14
  81   1   211  70   0   0   0      0.12     -0.03      1.28      1.30      0.14  0  0 2  13
  82   2  -213  53   0   0   0     -0.20      0.21     15.88     15.90      0.77  0  0 0  34
  83   2   213  53   0   0   0      0.39     -0.23      4.51      4.60      0.77  0  0 0  32
  84   2   111  53   0  92  95     -0.49     -0.36      8.47      8.50      0.13  0  0 0  23
  85   1  2212  53   0   0   0     -0.09      0.11     34.73     34.74      0.94  0  0 2   1
  86   1  -211  82   0   0   0     -0.42      0.07     13.24     13.25      0.14  0  0 2  14
  87   1   111  82   0   0   0      0.22      0.14      2.94      2.96      0.13  0  0 2  23
  88   2   211  83   0  96  97      0.39      0.02      1.25      1.31      0.14  0  0 0  13
  89   2   111  83   0  98 101      0.01     -0.26      2.44      2.46      0.13  0  0 0  23
\end{verbatim}
Next we find the entries from the formation zone cascade in
subroutine FOZOKL. 
The status of the initial state target nucleon is changed to ISTHKK=18.

In some of the entries we find the status code ISTHKK($i$)=16.
This are nucleons from secondary interactions not able to escape
from the nucleus. These nucleons are available as targets for
subsequent collisions.
\begin{verbatim}
  90   1   211  77  26   0   0      0.95      1.04      1.96      2.42      0.14  0  0 0  13
  91   1  2112  77  26   0   0     -0.69     -0.87      1.31      1.96      0.94  0  0 0   8
  92   2   111  84  14 102 103      0.11      0.12      0.12      0.24      0.13  0  0 0  23
  93   1  2112  84  14   0   0     -0.05     -0.06      0.07      0.95      0.94  0  0 0   8
  94   1  -211  84  14   0   0     -0.62     -0.37      7.82      7.86      0.14  0  0 0  14
  95   1   211  84  14   0   0      0.04      0.00      0.32      0.36      0.14  0  0 0  13
  96   1   211  88  18   0   0      0.79      0.01      0.91      1.22      0.14  0  0 0  13
  97   2  2212  88  18 104 105     -0.33     -0.04      0.30      1.04      0.94  0  0 0   1
  98   1  -211  89  31   0   0      0.64     -0.10      0.56      0.87      0.14  0  0 0  14
  99   1   211  89  31   0   0     -0.15     -0.11     -0.04      0.24      0.14  0  0 0  13
 100   1  2112  89  31   0   0     -0.25      0.01      0.64      1.16      0.94  0  0 0   8
 101   1   211  89  31   0   0     -0.19     -0.09      1.07      1.10      0.14  0  0 0  13
 102   1   111  92  19   0   0     -0.07      0.10     -0.09      0.20      0.14  0  0 0  23
 103   1  2112  92  19   0   0      0.19      0.07      0.08      0.96      0.94  0  0 0   8
 104  16  2212  97  22   0   0     -0.11     -0.10      0.03      0.95      0.94  0  0 0   1
 105   2  2112  97  22 106 107     -0.32      0.10      0.28      1.04      0.94  0  0 0   8
 106  16  2112 105  28 108   0     -0.05      0.08      0.21      0.97      0.94  0  0 0   8
 107   1  2212 105  28   0   0     -0.29     -0.10      0.11      0.99      0.94  0  0 0   1
\end{verbatim}
The final entries in the COMMON block refer to the evaporation
step of the excited residual target nucleus. With status code
ISTHKK($i$)=1000 we find the excited residual nucleus. 
 Evaporation nucleons, deexcitation photons and nuclear
 fragments follow with status
code ISTHKK($i$)=-1 and the stable residual  nuclei 
 follows with status code
ISTHKK($i$)=1001.
\begin{verbatim}
 1081000 80000   3 106 109 119     -0.09      0.32      0.67     23.43     23.42 25 14 0   0
 109  -1  2112 108   0   0   0     -0.01     -0.03     -0.08      0.94      0.94  0  0 2   8
 110  -1  2212 108   0   0   0     -0.07     -0.02      0.04      0.94      0.94  0  0 2   1
 111  -1  2212 108   0   0   0      0.07      0.15     -0.10      0.96      0.94  0  0 2   1
 112  -1  2212 108   0   0   0     -0.08      0.05     -0.02      0.94      0.94  0  0 2   1
 113  -1  2212 108   0   0   0      0.01      0.05      0.03      0.94      0.94  0  0 2   1
 114  -1  2212 108   0   0   0      0.04      0.01      0.05      0.94      0.94  0  0 2   1
 115  -1    22 108   0   0   0      0.00      0.00      0.00      0.00      0.00  0  0 2   7
 116  -1 80000 108   0   0   0     -0.09     -0.23      0.10      3.74      3.73  4  2 2   0
 117  -1 80000 108   0   0   0      0.06      0.09      0.27      3.74      3.73  4  2 2   0
 118  -1 80000 108   0   0   0      0.16      0.13      0.04      3.73      3.73  4  2 2   0
 1191001 80000 108   0   0   0     -0.18      0.13      0.33      6.55      6.54  7  3 2   0
\end{verbatim}

Eventually the actual final state hadrons, photons and nuclei 
are obtained
from the common block /HKKEVT/ by picking up all particles
with the status ISTHKK=1, ISTHKK=-1 and ISTHKK=1001.

%
%
%
\section*{Appendix C~: \\ Further options for internal use}
%
%
%
%
%
\begin{itemize}
   \item   Code word = 'PAULI'
     \begin{quote}
       This option monitors the inclusion of Pauli's principle 
       for the intranuclear cascade.  
     \begin{description}
      \item [ WHAT(1) :]  Pauli principle active for $1.0$ (default) \\
      \item [ WHAT(2) :]  triggers test prints for debugging
                                     if greater than $ 0.0$;\\
       default : $0.0$, i.e. no additional printout. 
      \end{description}
     \end{quote}
   \item   Code word = 'INTPT'
     \begin{quote}
       This option monitors the additional sampling of transverse 
       momenta for partons;  
     \begin{description}
      \item [ WHAT(1) :]  $1.0$ activates the sampling of
                      parton $p_t$~values (default). 
      \end{description}
     \end{quote}
   \item   Code word = 'TECALBAM'
     \begin{quote}
       This option triggers tests of the hadronization routines;
       additional input cards are required (see subroutine TECALB),
       to be given immediately after the code word definition;
       WHAT parameters have no meaning. 
     \end{quote}
   \item   Code word = 'HADRIN'
     \begin{quote}
     Via this option the secondary interactions generated from FHAD
       may be forced to be all inelastic/elastic ones. 
     \begin{description}
      \item [ WHAT(1) =] 0~:~inelastic/elastic as monitored by FOZOKL \\
                  1~:~inelastic collisions forced \\
                  2~:~elastic collisions forced.
      \end{description}
     \end{quote}
   \item   Code word = 'OUTLEVEL'
     \begin{quote}
	This option monitors the printout of intermediate results for
	diagnostics;
	\begin{description}
	   \item[WHAT(I) $\ge$ 1:]
	      diagnostics for remaining minor problems is printed
	      (mainly consistency/
	      accuracy problems in kinematical calculations).
	\end{description}
     \end{quote}
   \item   Code word = 'INFOREJE'
     \begin{quote}
	This option monitors the printout of rejection information
	\begin{description}
	   \item[WHAT(I) = 1:]
	   rejection information is printed.
	\end{description}
     \end{quote}
   \item   Code word = 'RECOMBIN'
     \begin{quote}
    Combinatoric transformation of s--s and v--v chains into s--v
and v--s chains
	\begin{description}
	   \item [WHAT(1)]  = 0 no recombination,\\
                 What(1) = 1 recombination implemented.\\         
                Default: 0. 
	\end{description}
     \end{quote}
   \item   Code word = 'CMHISTO'
     \begin{quote}
     The events in COMMON HKKEVT are sampled in the CMS frame
	\begin{description}
	   \item [WHAT(1)]  = 0 sampling in lab frame,\\
                 What(1) = 1 sampling in CMS frame\\         
                Default: 0. 
	\end{description}
	Note: If WHAT(1)=1 there is no possibility to treat the
	formation zone cascade. See cards TAUFOR and PROJKASK
     \end{quote}
   \item   Code word = 'FLUCTUAT'
     \begin{quote}
   implement cross section fluctuation (see Ref.
\cite{FLUCTUATION,FLUCTUATION1,FLUCTUATION2}) 
	\begin{description}
	   \item WHAT(1)  = 0 no fluctuations,\\
                 What(1) = 1 fluctuations implemented.\\         
                 Default: 0.  
	\end{description}
     \end{quote}
\end{itemize}
%
%
\section*{Appendix D~: \\ Further options from the DTUJET--99 code
within DPMJET}
%
%
%
%
%

\begin{itemize}
   \item   Code word:  STRUCFUN
     \begin{quote}
Defines the structure functions used in the sampling of hard constituent
scattering \cite{KMRS90,MRS92,CTEQ93} .
     \begin{description}
     \item [WHAT(1):] = ISTRUF \\
     ISTRUF =  ~21~: \hskip 0.4in Glueck,Reya,Vogt with K = 1:
GRV94LO\\ 
     ISTRUF =  ~22~: \hskip 0.4in Glueck,Reya,Vogt with K = 2:
GRV98LO\\ 
     ISTRUF =  ~23~: \hskip 0.4in CTEQ Collaboration (1996):
CTEQ96\\ 
     ISTRUF = ~221,222,223~: \hskip 0.4in as above with energy
dependent $p_{\perp}$ threshold value \\
The default is 222 . Only for this structure function we have a new fit.
Therefore the use of STRUCFUN with ISTRUF different from 222 
is not recommended in DPMJET--II.5.
     \end{description}

 An energy dependent cutoff\cite{DTUJET93} 
avoids a hard scattering cross section too large to be
treated in our simple eikonal approximation. To use this new
option the number 200 has to be added to the chosen ISTRUF value.
     \end{quote}

   \item   Code word:  PSHOWER

     \begin{quote}
This card determines whether hard partons initiate showers 
in connection with JETSET fragmentation.
     \begin{description}
     \item [WHAT(1):] = IPSHOW \\
   IPSHOW       =  ~0~:
 Generation of hard parton showers suppressed.\\
IPSHOW  =~1~:Hard parton showers are  included. \\
The default is { 1}. 
     \end{description}
     \end{quote}

\item Code word: GLUSPLIT
     \begin{quote}
Prevents splitting of gluons into quark--antiquark pairs.
     \begin{description}
     \item [WHAT(1):] = NUGLUU, Default: 1 \\
 NUGLUU=1~:Only one jet in hard gluon scattering.\\
Recommended: NUGLUU=1.
     \item [WHAT(2):] = NSGLUU, Default: 0 \\
NSGLUU=0~:Two jets in soft sea  gluon jets. \\
NSGLUU=1~:Only one jet in soft sea  gluon jets. \\
Recommended: NSGLUU=0.
     \end{description}
     \end{quote}
 
%
 
   \item   Code word:  SIGMAPOM
 
     \begin{quote}
Defines options and/or demands a test run for the calculation
of the DTU model unitarizing the soft and hard hadronic cross
sections.  The test run is for calculating the total and
inelastic cross sections as functions of the collision energy
as well as initializing and testing the sampling of multi-Pomeron
events at some typical energies.  Without the test run only the
initialization at the energy defined in the run is done.
     \begin{description}
     \item [WHAT(1):] = ITEST, test run for ITEST = 1.
\item [WHAT(2) =]~ ISIG, characterizes the soft and hard input cross
sections for the unitarization (see SUBROUTINE SIGSHD).\\
Recommended and default and only remaining option in DPMJET--II.5
:   ISIG = 10
\item [WHAT(3) =]~ IPIM characterizes the method to calculate
the distribution in the number of soft $n_s$ and hard $n_h$ Pomerons\\
Default and recommended and only remaining option in DPMJET--II.5:
IPIM = 482
\item [WHAT(4)~=]~ LMAX Default: 30\\
 LMAX~:Maximum considered number of soft Pomerons (Less
than 26).
\item [WHAT(5)~=]~ MMAX Default: 100\\
 MMAX~:Maximum considered number of hard Pomerons
(Less than 101).
\item [WHAT(6)~=]~ NMAX Default: 2\\
 NMAX~:Maximum considered number of triple--Pomerons
 (Less than 13).
     \end{description}
     \end{quote}

\end{itemize}

%
\end{document}